\documentclass[useAMS,usenatbib]{mn2e}

\usepackage{graphicx}
\usepackage{txfonts}

\def\Chan{{\sl Chandra}}
\def\XMM{XMM-{\sl Newton}}
\def\COS{HST-COS}
\def\STIS{STIS}
\def\Lya{Ly\,$\alpha$}
\def\CIV{C\,{\sc iv}}
\def\CV{C\,{\sc v}}
\def\CVI{C\,{\sc vi}}
\def\NIV{N\,{\sc iv}]}
\def\NV{N\,{\sc v}}
\def\NVI{N\,{\sc vi}}
\def\NVII{N\,{\sc vii}}
\def\OV{O\,{\sc v}}
\def\OVI{O\,{\sc vi}}
\def\OVII{O\,{\sc vii}}
\def\OVIII{O\,{\sc viii}}
\def\NeVII{Ne\,{\sc vii}}
\def\NeVIII{Ne\,{\sc viii}}
\def\NeIX{Ne\,{\sc ix}}
\def\NeX{Ne\,{\sc x}}
\def\MgIX{Mg\,{\sc ix}}
\def\MgX{Mg\,{\sc x}}
\def\MgXI{Mg\,{\sc xi}}
\def\MgXII{Mg\,{\sc xii}}

\def\SiII{Si\,{\sc ii}}
\def\SiIV{Si\,{\sc iv}}

\def\SiXI{Si\,{\sc xi}}
\def\SiXII{Si\,{\sc xii}}
\def\SiXIII{Si\,{\sc xiii}}
\def\SX{S\,{\sc x}}
\def\SXI{S\,{\sc xi}}
\def\SXII{S\,{\sc xii}}
\def\SXIII{S\,{\sc xiii}}

\def\FeIX{Fe\,{\sc ix}}

\def\FeXVI{Fe\,{\sc xvi}}
\def\FeXVII{Fe\,{\sc xvii}}
\def\FeXVIII{Fe\,{\sc xviii}}
\def\FeXIX{Fe\,{\sc xix}}
\def\FeXX{Fe\,{\sc xx}}
\def\FeXXI{Fe\,{\sc xxi}}
\def\FeXXII{Fe\,{\sc xxii}}

\title[The X-ray/UV absorber in NGC 4593]{The X-ray/UV absorber in NGC 4593}
\author[J. Ebrero et al.]{J. Ebrero$^{1}$\thanks{E-mail:
J.Ebrero.Carrero@sron.nl}, J. S. Kaastra$^{1}$, G. A. Kriss$^{2,3}$, C. P. de Vries$^{1}$, and E. Costantini$^{1}$ \\
$^{1}$SRON - Netherlands Institute for Space Research, Sorbonnelaan 2, 3584 CA, Utrecht, The Netherlands\\
$^{2}$Space Telescope Science Institute, 3700 San Martin Drive, Baltimore, MD 21218, USA\\
$^{3}$Department of Physics and Astronomy, The Johns Hopkins University, Baltimore, MD 21218, USA}
\begin{document}

\date{Accepted <date>. Received <date>; in original form <date>}

\pagerange{\pageref{firstpage}--\pageref{lastpage}} \pubyear{2013}

\maketitle

\label{firstpage}

\begin{abstract}
We present the results of a recent (March 2011) 160~ks \Chan{}-LETGS observation of the Seyfert galaxy NGC 4593, and the analysis of archival X-ray and UV spectra taken with \XMM{} and HST/STIS in 2002. We find evidence of a multi-component warm absorber (WA) in the X-rays with four distinct ionisation degrees ($\log \xi = 1.0$, $\log \xi = 1.7$, $\log \xi = 2.4$, and $\log \xi = 3.0$) outflowing at several hundreds of km s$^{-1}$. In the UV we detect 15 kinematic components in the absorbers, blueshifted with respect to the systemic velocity of the source, ranging from $-60$~km s$^{-1}$ to $-1520$~km s$^{-1}$. Although the predicted \CIV{} and \NV{} column densities from the low-ionisation X-ray outflow are in agreement with those measured for some components in the STIS spectrum, there are kinematic discrepancies that may prevent both the X-ray and UV absorbers from originating in the same intervening gas. We derive upper limits on the location of the absorbers finding that the high-ionisation gas lie within $\sim 6 - 29$~pc from the central ionising source, while the low-ionisation gas is located at several hundreds of pc. This is consistent with our line of sight passing through different parts of a stratified wind. The total kinetic energy of the outflows injected into the surroundings of the host galaxy only accounts for a tiny fraction of the bolometric luminosity of the source and it is therefore unlikely that they may cause a significant impact in the interstellar medium of NGC 4593 in a given single episode of activity.
\end{abstract}


\section{Introduction}
\label{intro}

Active galactic nuclei (AGN) are powered by accretion of matter onto the central supermassive black hole (SMBH) that is believed to reside in their centre (see \citealt{Rees84} for a review). The large bolometric output of AGN is powerful enough to drive ionised winds into the medium of the host galaxy in the form of mass outflows. Indeed, more than 50\% of nearby Seyfert 1 galaxies present clear evidence of photoionised gas, the so-called warm absorber (WA hereafter), in their soft X-ray spectra (e.g. \citealt{Rey97}; \citealt{Geo98}; \citealt{CKG03}), and in their UV spectra (\citealt{Cre99}; \citealt{Kriss02}; \citealt{Dunn10}).

These WA are usually detected as absorption features blushifted with respect to the systemic velocity of the source, typically by hundreds of km s$^{-1}$ (\citealt{Kaa00}; \citealt{Kas01}). These mass outflows inject mass and energy into the interstellar medium (ISM) and can deeply affect the evolution of the host galaxy (\citealt{SR98}; \citealt{WL03}) and even the surrounding intergalactic medium (IGM, \citealt{Gra04}; \citealt{SO04}). If the kinetic luminosity of the outflows is as high as 0.5\% to 5\% of the bolometric luminosity (\citealt{HE10}; \citealt{DM05}), their impact is enough to significantly regulate galaxy growth and reproduce the observed $M-\sigma$ relation (\citealt{FM00}; \citealt{Geb00}).

In spite of their importance, the structure, origin, launching mechanism, and relation between the X-ray and UV absorbers are not yet fully understood. The proposed origins for AGN outflows are disc-driven winds (\citealt{MC95}; \citealt{Elvis00}) and thermal winds produced by evaporation from the torus (\citealt{KK01}). Winds emanating from the accretion disc are expected to lie within several hundred gravitational radii, while thermally-driven winds should originate at pc scales, at or beyond the location of the dusty torus. An accurate determination of the distance is essential to establish the actual mass outflow rate and assess their importance as contributors for cosmic feedback processes.

In this paper we present the results of the analysis of a $160$~ks observation with \Chan{}-LETGS and archival X-ray and UV observations with \XMM{} and HST-STIS, respectively, of the nearby ($z = 0.0090 \pm 0.0001$, \citealt{Str92}) Seyfert 1 galaxy NGC 4593. It is a mildly luminous AGN ($L_{\rm 2-10} \sim 5 \times 10^{42}$~erg s$^{-1}$, this work) and is hosted by a barred spiral galaxy with a relatively modest-size SMBH in its centre ($M_{\rm BH} = 9.8 \pm 2.1 \times 10^6$~$M_\odot$, \citealt{Den06}). NGC 4593 has been subject to a number of prior observations in the X-ray domain. \citet{Steen03} found evidence of an ionised absorber at the redshift of the source using \Chan{}-LETGS and \XMM{} observations. \citet{McK03} confirmed the presence of this WA using a \Chan{}-HETGS observation, and adequately described the observed features using a single-zone photoionised gas, although some features remained unidentified. \citet{Bre07} found evidence of a multi-zone layer of absorbing material intrinsic to the source using a 76~ks \XMM{} EPIC-pn observation. They could not further constrain its physical parameters as the RGS spectrum in this observation was not used.

This paper is organised as follows. In Sect.~\ref{data} we present the X-ray and UV data and the reduction techniques used. In Sect.~\ref{analysis} we describe the spectral analysis and the different models used to characterize the X-ray and UV spectra. In Sect.~\ref{discussion} we discuss our results and our conclusions are summarised in Sect.~\ref{conclusions}. Throughout this paper we assume a cosmological framework with $H_{\rm 0} = 70$~km s$^{-1}$~Mpc$^{-1}$, $\Omega_{\rm M} = 0.3$, and $\Omega_\Lambda = 0.7$. The quoted errors refer to 68.3\% confidence level ($\Delta C = 1$ for one parameter of interest) unless otherwise stated.

\section{The data}
\label{data}

\begin{figure}
  \includegraphics[width=6.5cm,angle=90]{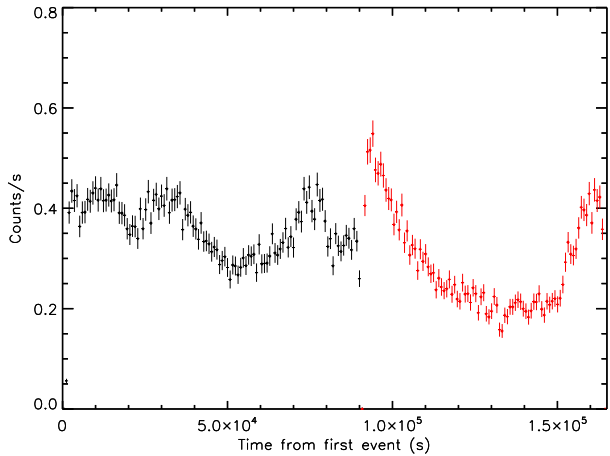}
  \caption{\Chan{}~lightcurve of NGC 4593. Time bin size is 100~s.}
  \label{lcchandra}
\end{figure}

NGC 4593 was observed on two consecutive orbits in March 2011 by the Low Energy Transmission Grating Spectrometer (LETGS, \citealt{Bri00}) with the HRC-S detector on board of \Chan{} for a total exposure time of $\sim$160~ks. The LETGS data were reduced using the standard pipeline (CIAO 4.2) until the level 1.5 event files were created. The rest of the procedure, until the creation of the final level 2 event file, followed the same steps as the standard pipeline regarding the wavelength accuracy determination and effective area generation, but was carried out using an independent procedure first described in \citet{Kaa02}.

No episodes of short-term variability (e.g. within a few~ks) were observed after applying a bin size of 100~s. During the first observation the source flux varies smoothly, while during the second observation the source experiences a drop in flux of the order of $\sim$50\% followed by a recovery in the last part of the observation (see Fig.~\ref{lcchandra}). Because of the low statistics, no significant spectral variations were observed between the high and low spectral states (i.e. spectral fits for the low and high flux states rendered the same parameter values within the uncertainties). The combined spectra were therefore used in the analysis described throughout this paper.

We also retrieved a \XMM{} observation of NGC 4593 taken in 2002 from the \XMM{} Science Archive (XSA). The analysis of this observation was first reported in \citet{Bre07} but they did not include the data from the Reflection Grating Spectrometer (RGS, \citealt{Her01}). We processed these data with the standard \XMM{} analysis system SAS version 11. Only data from the second, restarted exposure U002 were used, since the first (default) exposure (S004/S005) was stopped almost immediately due to high radiation and contained virtually no data. We used cross dispersion and CCD pulse-height selections of 95\%, and background exclusion of 98\% (the SAS default settings). The spectra were binned in beta space and the response matrix used 9\,000 bins. The net exposure time of this observation was about 77~ks. The \Chan{} LETGS and \XMM{} observation logs are shown in Table~\ref{letgslog}.

In the UV, NGC 4593 was observed with the Space Telescope Imaging Spectrograph (STIS) in June 2002 for $\sim$11\,000~s, using the echelle E140M grating and the $0.2 \times 0.2$ arcsec aperture. The echelle spectrum was directly retrieved from the MAST archive with all up-to-date calibrations applied, including corrections for scattered light in the echelle mode (\citealt{Val02}) and for the echelle blaze corrections (\citealt{Alo07}). The STIS observation log of NGC 4593 is shown in Table~\ref{stislog}.

\begin{table}
  \caption{NGC 4593 X-ray observation log. Column (1): facility name; column (2): observation ID number; column (3): date of the observation; column (4): GMT start time of the observation; column (5): exposure time of the observation.}
  \label{letgslog}
  \begin{tabular}{@{}lcccc}
    \hline
    Observatory & Obs ID & Date & Start Time & Exp. Time \\
                &        & (dd-mm-yyyy) & (hh:mm:ss) & (s) \\
    (1)         & (2)    & (3)          & (4)   & (5) \\
    \hline
    \Chan{} & 12154 & 02-03-2011 & 19:59:04 & 88\,730 \\
    \Chan{} & 13240 & 05-03-2011 & 03:58:07 & 72\,630 \\
    \hline
    \XMM{} & 0059830101 & 23-06-2002 & 10:05:05 & 87\,238 \\
    \hline
  \end{tabular}
\end{table}

\begin{table}
  \caption{NGC 4593 STIS observation log. Column (1): Dataset label; column (2): grating used in the observation; column (3): date of the observation; column (4): GMT start time of the observation; column (5): exposure time of the observation.}
  \label{stislog}
  \begin{tabular}{@{}lcccc}
    \hline
    Data Set & Grating & Date & Start Time & Exp. Time \\
             &         & (dd-mm-yyyy) & (hh:mm:ss) & (s) \\
    (1)      & (2)     & (3)          & (4)   & (5) \\
    \hline
    O5L501020 & E140M & 23-06-2002 & 22:30:02 & 2\,287 \\
    O5L501030 & E140M & 23-06-2002 & 23:55:09 & 2\,899 \\
    O5L501040 & E140M & 24-06-2002 & 01:31:19 & 2\,899 \\
    O5L501050 & E140M & 24-06-2002 & 03:07:28 & 2\,899 \\
    \hline
  \end{tabular}
\end{table}

\begin{figure*}
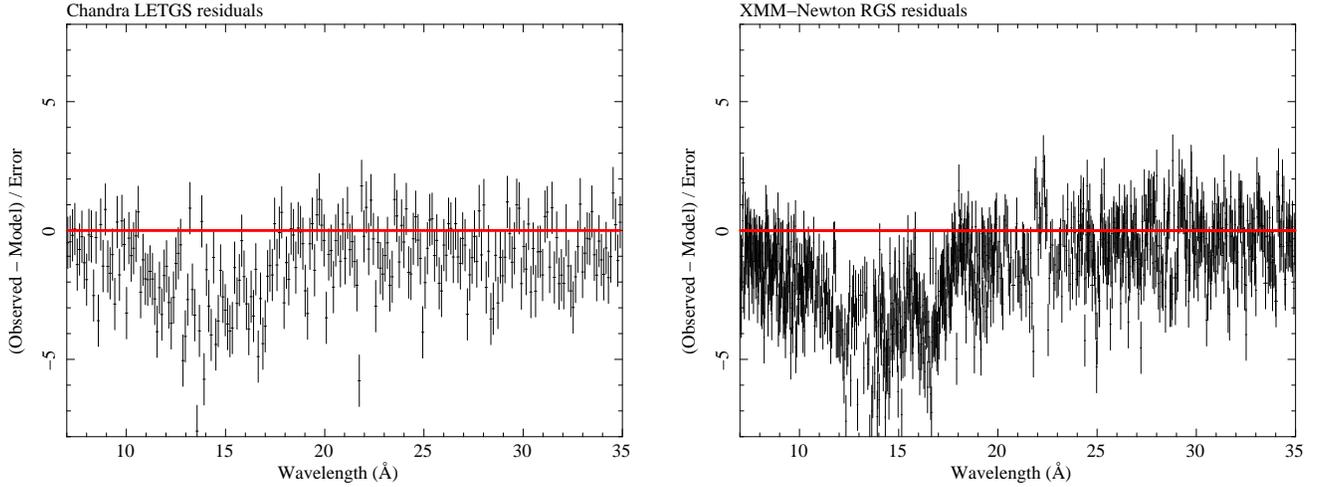

  \hbox{
    \includegraphics[width=6.5cm,angle=-90]{letgs_res.ps}
    \includegraphics[width=6.5cm,angle=-90]{rgs_res.ps}
  }
  \caption{Residuals of the \Chan{}~LETGS (left panel) and \XMM{}~RGS (right panel) spectra after continuum fitting (power-law plus modified black body emission). Emission lines and absorption models are not included. The wavelength scales are in the observer frame.}
  \label{residuals}
\end{figure*}

\begin{table}
  \caption{LETGS best-fit continuum and emission lines parameters. Column (1): model used; column (2): photon index; column (3): black body temperature; column (4): $0.5-2$~keV flux of the model; column (5): emission line; column (6): rest wavelength of the line; column (7): full width at half maximum of the line; column (8): line flux.}
  \label{cont}
  \begin{tabular}{@{}lccc}
    \hline
    Model  &  $\Gamma$  & $T$  &  $F_{0.5-2~ \rm keV}$ \\
           &            & (keV) &   (erg~cm$^{-2}$~s$^{-1}$) \\
    (1)    &  (2)       & (3)   &  (4)  \\
    \hline
    Power law & $1.81 \pm 0.03$ & $\dots$ & $1.22 \pm 0.02 \times 10^{-11}$ \\
    Mod. black body & $\dots$  &  $0.12 \pm 0.01$ & $1.11 \pm 0.14 \times 10^{-12}$ \\
    \hline
    Line & Wavelength & $FWHM$ & $F_{\rm line}$ \\
         &   (\AA)      &  (\AA) &  ($\times 10^{-14}$~erg~cm$^{-2}$~s$^{-1}$) \\
    (5)  &  (6)        & (7)     &  (8)   \\
    \hline
    \NeIX{} (f) & 13.698 & $0.09 \pm 0.05$ & $1.8 \pm 0.8$ \\
    \OVIII{}    & 18.969 & $< 0.10$ & $1.5 \pm 1.3$ \\
    \OVII{} (i) & 21.808 & $< 0.09$ & $1.1 \pm 0.3$ \\
    \OVII{} (f) & 22.101 & $0.05 \pm 0.03$ & $2.0 \pm 0.6$ \\
    \NVI{} (i)  & 29.083 & $< 0.13$ & $0.7 \pm 0.5$ \\
    \NVI{} (f)  & 29.535 & $< 0.30$ & $0.9 \pm 0.4$ \\
    \CVI{}      & 33.736 & $< 0.08$ & $0.6 \pm 0.4$ \\
    \hline
  \end{tabular}
\end{table}

\begin{table}
  \caption{RGS best-fit continuum and emission lines parameters. Column (1): model used; column (2): photon index; column (3): black body temperature; column (4): $0.5-2$~keV flux of the model; column (5): emission line; column (6): rest wavelength of the line; column (7): full width at half maximum of the line; column (8): line flux.}
  \label{rgscont}
  \begin{tabular}{@{}lccc}
    \hline
    Model  &  $\Gamma$  & $T$  &  $F_{0.5-2~ \rm keV}$ \\
           &            & (keV) &   (erg~cm$^{-2}$~s$^{-1}$) \\
    (1)    &  (2)       & (3)   &  (4)  \\
    \hline
    Power law & $1.98 \pm 0.03$ & $\dots$ & $2.4 \pm 0.2 \times 10^{-11}$ \\
    Mod. black body & $\dots$  &  $0.13 \pm 0.01$ & $2.2 \pm 0.2 \times 10^{-12}$ \\
    \hline
    Line & Wavelength & $FWHM$ & $F_{\rm line}$ \\
         &   (\AA)      &  (\AA) &  ($\times 10^{-14}$~erg~cm$^{-2}$~s$^{-1}$) \\
    (5)  &  (6)        & (7)     &  (8)   \\
    \hline
    \NeIX{} (f) & 13.698 & $< 0.18$ & $1.8 \pm 0.9$ \\
    \OVIII{}    & 18.969 & $< 0.10$ & $2.7 \pm 1.0$ \\
    \OVII{} (i) & 21.808 & $< 0.11$ & $4.0 \pm 1.0$ \\
    \OVII{} (f) & 22.101 & $0.18 \pm 0.07$ & $7.0 \pm 1.6$ \\
    \NVI{} (i)  & 29.083 & $0.22 \pm 0.06$ & $4.6 \pm 1.0$ \\
    \NVI{} (f)  & 29.535 & $< 0.06$ & $1.9 \pm 0.6$ \\
    \CVI{}      & 33.736 & $0.16 \pm 0.06$ & $2.5 \pm 1.0$ \\
    \hline
  \end{tabular}
\end{table}

\section{Analysis}
\label{analysis}

The combined LETGS spectrum, and the \XMM{} RGS spectrum as well as the STIS spectrum were analysed using the {\it SPEX}\footnote{{\tt http://www.sron.nl/spex}} fitting package v2.04 (\citealt{Kaa96}). The adopted fitting method was C-statistics (\citealt{Cash79}) so that binning of the data is in principle no longer needed. However, the data were rebinned by a factor of 3 to avoid oversampling. In this section we describe the different models and techniques used to model the continuum and the absoption and emission features present in the spectra of NGC 4593. The adopted cosmological redshift of NGC 4593 is $z=0.009$ (\citealt{Str92}).

\subsection{X-rays}
\label{xanal}

In this section we report the X-ray analysis of the \Chan{}~and \XMM{}~observations. We first model the continuum and emission lines in the LETGS and RGS spectra, while the absorption features are described using two independent models: a first approach using the {\it slab} model in {\it SPEX}, and a more physical photoionization model using {\it xabs}.

\begin{figure*}
  \hbox{
    \includegraphics[width=6.5cm,angle=-90]{letgs_fluxed_520.ps}
    \includegraphics[width=6.5cm,angle=-90]{rgs_fluxed_616.ps}
  }
  \hbox{
    \includegraphics[width=6.5cm,angle=-90]{letgs_fluxed_2035.ps}
    \includegraphics[width=6.5cm,angle=-90]{rgs_fluxed_1626.ps}
  }
  \hbox{
    \includegraphics[width=6.5cm,angle=-90]{letgs_fluxed_3550.ps}
    \includegraphics[width=6.5cm,angle=-90]{rgs_fluxed_2637.ps}
  }
  \caption{\Chan{} LETGS spectrum (left panels) and \XMM{} RGS spectrum (right panels) of NGC 4593. RGS1 and RGS2 are plotted together. The solid line represent the best-fit model with {\it slab}. Some of the most relevant features are labeled. The wavelength scales are in the observer frame.}
  \label{Xspec}
\end{figure*}

\begin{figure*}
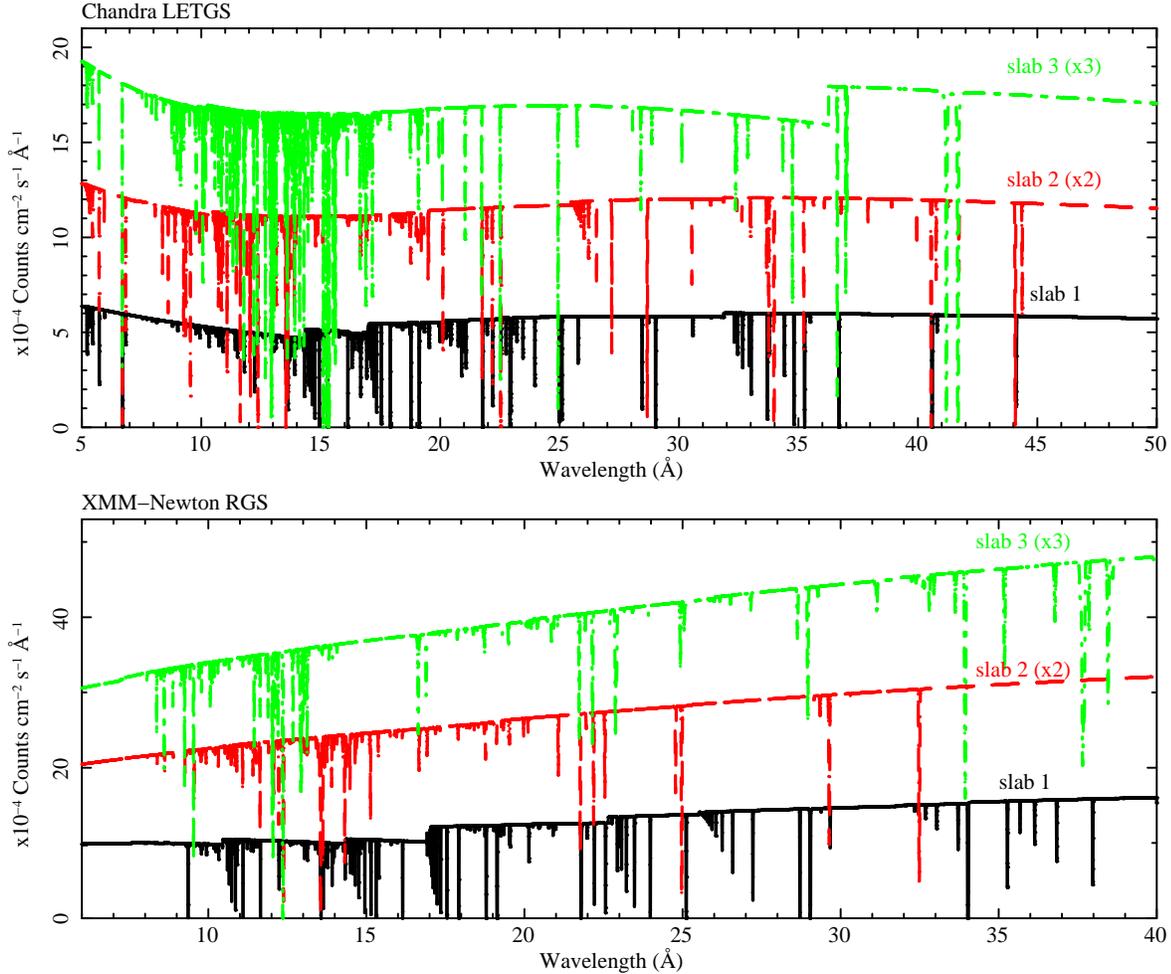

  \hbox{
    \includegraphics[width=6.5cm,angle=-90]{letgs_slab.ps}
  }
  \hbox{
    \includegraphics[width=6.5cm,angle=-90]{rgs_slab.ps}
  }
  \caption{Model plot showing the contribution of each of the {\it slab} components to the \Chan{}~LETGS (upper panel) and the \XMM{}~RGS (lower panel) spectra. In both cases the solid line represents {\it slab} component 1, while the dashed and dot-dashed lines represent {\it slab} components 2 and 3, respectively. For clarity, {\it slab} components 2 and 3 have been shifted in flux by a factor 2 and 3, respectively. The wavelength scales are in the observer frame.}
  \label{slabmodel}
\end{figure*}

\subsubsection{Continuum and emission lines}
\label{xraycont}

A visual inspection of the combined LETGS spectrum reveals a wealth of absorption lines, including a broad dip in the continuum in the $10-18$~\AA~range (see Fig.~\ref{residuals}). Some significant emission lines are also present (see Fig.~\ref{Xspec}, left panels). The spectrum at wavelengths longer than 60~\AA~was dominated by the background so in what follows the spectral fits where carried out in the $5-60$~\AA~range.

The continuum was modeled with a power law. Local foreground absorption was set to $N_{\rm H} = 2.31 \times 10^{20}$~cm$^{-2}$ (\citealt{DL90}) using the {\it hot} model in {\it SPEX}, fixing the temperature to $0.5$~eV to mimic a neutral gas. The excess in soft X-rays was accounted for adding a modified black body model ({\it mbb} model in {\it SPEX}), which considers modifications of a simple black body model by coherent Compton scattering and it is based on the calculations of \citet{KB89}. The power law provides a photon index $\Gamma = 1.81 \pm 0.03$, and the modified black body has a temperature $T = 0.12 \pm 0.01$~keV. The best-fit parameters and fluxes of the LETGS continuum are reported in Table~\ref{cont}.

The emission features were modeled with Gaussian lines. We kept the centroid wavelength frozen to the laboratory values, while the normalization and line widths were left as free parameters. Emission lines of \OVIII{}~and \CVI{}~are marginally detected at the $\sim 1.5\sigma$ confidence level. Some lines of the He-like triplets of \NeIX{}, \OVII{}, and \NVI{}~are also detected although not all of them, and with variable significance. The forbidden ($f$) lines of \NeIX{}~and \OVII{} are significantly detected at the $3\sigma$ confidence level, while that of \NVI{} is only detected at the $\sim 2\sigma$ level. Likewise, the intercombination ($i$) line of \OVII{}~is significantly detected above the $3\sigma$ confidence level, while \NVI{}~(i) is marginally detected at the $\sim 1\sigma$ level, and \NeIX{}~(i) is statiscally undetected. The resonance ($r$) lines of these species are statistically undetected in this dataset, even though the low continuum flux favours the detection of emission lines, which indicates that the main ionisation process is photoionisation (\citealt{PD00}). The best-fit parameters of the emission lines in the LETGS spectrum are reported in Table~\ref{cont}.

The RGS spectrum of NGC 4593 was analysed in the same manner as the LETGS combined spectrum. The fitting range was set to the $5 - 37$~\AA~interval. The local foreground absorption was set to the same parameter values and the continuum was also fitted using a simple power law plus a modified black body model. The overall flux of the source in the $0.5-2$~keV range at the time of the \XMM{}~observation was higher than that of \Chan{} by almost a factor of 2, and the power law slope was steeper ($\Gamma = 1.98 \pm 0.03$). The normalization of the modified black body was also higher by a factor of 2, although the temperature remained the same as in the LETGS spectrum ($T = 0.13 \pm 0.01$). The best-fit values are reported in Table~\ref{rgscont}.

Despite the higher continuum level in the RGS spectrum with respect to that of the LETGS one, we were also able to detect several emission features with significances at the $\sim 3\sigma$ confidence level or better, thanks to the better signal-to-noise ratio of the spectrum. The lines were modeled using again Gaussians with the centroid wavelength fixed, and the normalization and width of the line as free parameters. The species detected were the same as in the LETGS spectrum, and their best-fit parameters are reported in Table~\ref{rgscont}. We further discuss the emission lines in NGC 4593 in Sect.~\ref{emission}.

\begin{table}
  \caption{Absorption lines identifications in the LETGS and RGS spectra of NGC 4593. Column (1): ion; column (2): transition; column (3): laboratory wavelength; column (4): statistical significance of the line in the LETGS spectrum; column (5): statistical significance of the line in the RGS spectrum.}
  \label{abslines}
  \begin{tabular}{@{}llccc}
    \hline
    Ion  &  Transition  &  $\lambda_{\rm 0}$  &  $\Delta {\rm C}^2$  &  $\Delta {\rm C}^2$  \\
         &              &  (\AA)          &      (LETGS)            &  (RGS)  \\
    (1)        &  (2)             & (3)          & (4)          & (5) \\
    \hline
    \CV{}  & $1s^2-1s2p\, ^1P_1$ & 40.268 & 3 & $\dots$ \\
           & $1s^2-1s3p\, ^1P_1$ & 34.973 & 4 & 6 \\
    \CVI{} & $1s-2p\, (\rm Ly\alpha)$ & 33.736 & 3 & 28 \\
           & $1s-3p\, (\rm Ly\beta)$ & 28.466 & 7 & 15 \\
           & $1s-4p\, (\rm Ly\gamma)$ & 26.990 & 4 & 16 \\
    \NVI{} & $1s^2-1s2p\, ^1P_1$ & 28.788 & 4 & 23 \\
           & $1s^2-1s3p\, ^1P_1$ & 24.900 & 7 & 14 \\
    \NVII{} & $1s-2p\, (\rm Ly\alpha)$ & 24.781 & 4 & 38 \\
           & $1s-3p\, (\rm Ly\beta)$ & 20.910 & 1 & 7 \\
    \OV{}  & $1s-2p$ & 22.370 & 6 & 16 \\
    \OVI{} & $1s^22s-1s2s(^1S)2p$ & 22.019 & 3 & 23 \\
    \OVII{} & $1s^2-1s2p\, ^1P_1$ & 21.602 & 15 & 25 \\
            & $1s^2-1s3p\, ^1P_1$ & 18.628 & 3 & 7 \\
            & $1s^2-1s4p\, ^1P_1$ & 17.768 & 3 & 6 \\
            & $1s^2-1s5p\, ^1P_1$ & 17.396 & 1 & 6 \\
            & $1s^2-1s6p\, ^1P_1$ & 17.200 & 2 & 6 \\
    \OVIII{} & $1s-2p\, (\rm Ly\alpha)$ & 18.969 & 6 & 11 \\
             & $1s-3p\, (\rm Ly\beta)$ & 16.006 & 6 & 35 \\
             & $1s-4p\, (\rm Ly\gamma)$ & 15.176 & 2 & 7 \\
             & $1s-5p\, (\rm Ly\delta)$ & 14.821 & 2 & 9 \\
    \NeVII{} & $1s-2p$ & 13.814 & 8 & 19 \\
    \NeVIII{} & $1s^22s-1s2s(^1S)2p$ & 13.653 & 2 & 7 \\
    \NeIX{} & $1s^2-1s2p\, ^1P_1$ & 13.447 & 23 & 45 \\
            & $1s^2-1s3p\, ^1P_1$ & 11.547 & 8 & 6 \\
            & $1s^2-1s4p\, ^1P_1$ & 11.000 & 2 & 6 \\
    \NeX{} & $1s-2p\, (\rm Ly\alpha)$ & 12.132 & 2 & 14 \\
           & $1s-3p\, (\rm Ly\beta)$ & 10.238 & 1 & 6 \\
    \MgIX{} & $2s^2-1s2s^22p$ & 9.378 & 3 & 6 \\
    \MgX{} & $1s^2(^1S)2s-1s(^2S)2s2p(^3P^0)$ & 9.281 & 3 & 6 \\
    \MgXI{} & $1s^2-1s2p\, ^1P_1$ & 9.169 & 2 & 10 \\
            & $1s^2-1s3p\, ^1P_1$ & 7.851 & 3 & 7 \\
    \MgXII{} & $1s-2p$ & 8.419 & 4 & 6 \\
    \SiXI{} & $2s-3p\, ^1P_1$ & 43.762 & 13 & $\dots$ \\
    \SiXII{} & $2s-3p\, ^2P_{1/2}$ & 40.911 & 3 & $\dots$ \\
    \SiXIII{} & $1s2p(^1P_1)-1s^2(^1S_0)$ & 6.648 & 15 & 8 \\
    \SX{} & $2s^22p^3-2s^22p^2(^3P)2d$ & 42.495 & 2 & $\dots$ \\
    \SXI{} & $2s^22p^2-2s^22p3d$ & 39.240 & 5 & $\dots$ \\
    \SXII{} & $2s^22p-2s^23d$ & 36.398 & 4 & 6 \\
    \SXIII{} & $2s^2-2s3p$ & 32.238 & 4 & 18 \\
    \hline
  \end{tabular}
\end{table}

\subsubsection{Absorption lines: modeling with {\it slab}}
\label{xrayslab}

The absoption features seen in the X-ray spectra are mostly attributed to the presence of ionised gas at the redshift of the source (\citealt{McK03}; \citealt{Steen03}). Before performing any global modeling of the spectra, we examined them to identify the different absorption troughs in a one-by-one basis. While this is possible to a considerable extent, some of identifications cannot be resolved in this way because of blending with a neighboring transition, the most clear example being the plethora of transitions in the Fe UTA. In Table~\ref{abslines} we report the majority of the absorption lines seen in the LETGS and RGS spectra, along with their statistical significance expressed in terms of $\Delta {\rm C}^2$ obtained by fitting a Gaussian profile to each of them. The absorption troughs are easier to detect in the RGS spectrum because of the better signal-to-noise with respect to the LETGS spectrum. To globally characterize the WA in NGC 4593, we firstly used a {\it slab} component in both \Chan{}~and \XMM{}~spectra. The {\it slab} model in {\it SPEX} calculates the transmission of a slab of material without any assumption on the underlying ionisation balance. The free parameters of this model are the ionic column densities, the {\it rms} velocity broadening, and the outflow velocity. 

The addition of a single {\it slab} component to the \Chan{}~LETGS spectrum significantly improved the fit from ${\rm C^2}/\nu = 2834/1494 \sim 1.9$ to ${\rm C^2}/\nu = 1850/1463 \sim 1.3$, although several troughs remained unfitted. Some of the lines seemed to belong to the same ionisation species albeit with slightly different outflow velocities. For this reason we employed a fitting strategy that included all ions for each velocity component in the same manner as \citet{Det11}. This model is physically more realistic as it allows a multi-velocity structure for every absorption line, consistent with what is typically found in the UV spectra of AGN. The addition of a second {\it slab} component further improves the fit by $\Delta{\rm C^2}/\Delta\nu = 167/31$. Visually, the fit was not good enough yet with several residuals at higher energies, where most of the higher ionisation species are expected to lie. Therefore, we added a third {\it slab} component that improved the fit by a further $\Delta{\rm C^2}/\Delta\nu = 161/31$ for a final ${\rm C^2}/\nu = 1522/1401 \sim 1.08$. The addition of more components no longer improved the fit. 

The basic WA structure in NGC 4593 depicted by the fit of the LETGS spectrum suggest the presence of a multi-kinematic absorption system with outflow velocities of $-270 \pm 40$, $-545 \pm 55$, and $-820 \pm 80$~km s$^{-1}$, respectively, with velocity widths $\sigma \sim 55$~km s$^{-1}$. The differences in the outflow velocities of the three components are not dramatic, all of them outflowing at modest velocities of a few hundreds of km s$^{-1}$, typical of the WA often seen in Seyfert 1 galaxies (\citealt{Blu05}). The ionic column densities obtained in the fit for the most relevant ions are reported in Table~\ref{slab}. From these results it becomes apparent that the faster component accounts for the majority of the higher ionisation species, while the slower component accounts for the less ionised ones.

A similar approach was followed to model the absorbers in the \XMM{}~RGS spectrum of NGC 4593. Compared with the LETGS spectrum, the RGS data has a higher flux and better signal-to-noise ratio, which favors the detection of absorbtion features. The addition of a single {\it slab} component renders a goodness of fit of ${\rm C^2}/\nu \sim 1.3$, but it is not visually satisfactory. A second {\it slab} component significantly improves the fit by $\Delta{\rm C^2}/\Delta\nu = 288/30$. Similarly to the LETGS fitting, we added a third {\it slab} component that accounted for the higher velocity ions, obtaining a final fit improvement by $\Delta{\rm C^2}/\Delta\nu = 97/30$, slightly below the $2\sigma$ level, for a final best-fit with ${\rm C^2}/\nu \sim 1.11$. A model plot of the different {\it slab} components in the LETGS and RGS spectra is shown in Fig.~\ref{slabmodel}.

The three kinematic components detected in the RGS spectrum have outflow velocities of $-70 \pm 40$, $-380 \pm 60$, and $-880 \pm 90$~km s$^{-1}$, respectively. While there seems to be an apparent mismatch between the velocities measured by \Chan{}~and \XMM{}, it is known that there is a systematic uncertainty in the RGS wavelength. Such uncertainty translates into a difference between the LETGS and RGS wavelength scales of $\sim 7$~m\AA~, which corresponds to a velocity offset of 110~km~s$^{-1}$ at a typical wavelength of 20~\AA. Therefore, there is {\it no discrepancy} between the RGS and LETGS wavelength scales within the systematic uncertainties. Moreover, in spite of the flux variation between the \XMM{}~and \Chan{}~observations (roughly a factor of 2 higher in the former), no dramatic changes in the ionic column densities of the different species are observed. These values are also reported in Table~\ref{slab}. Although the {\it slab} modeling makes no {\it a priori} assumption on the underlying ionising continuum, as a guide we also report in Table~\ref{slab} the value of the ionisation parameter $\log \xi$ (which is not a parameter of the {\it slab} model) at which each of the species reachs a peak in its concentration, as provided by the spectral energy distribution for each observation (see Sect.~\ref{xrayxabs}).

\begin{table*}
  \begin{minipage}{170mm}
  \caption{Ionic column densities of the most relevant ions in the NGC 4593 outflow. Column (1): Ion; columns (2)-(4): ionic column densities for the LETGS components; column (5): ionisation parameter at which the ion has its peak concentration given by the SED of the \Chan{}~observation; columns (6)-(8): ionic column densities for the RGS components; column (9): ionisation parameter at which the ion has its peak concentration given by the SED of the \XMM{}~observation.}
  \label{slab}
  \begin{tabular}{@{}lcccc|cccc}
    \hline
             &      &   \Chan{}~LETGS  &      &     &     &    \XMM{}~RGS &      &    \\
    \hline
    Component &     1      &      2   &   3    &     &    1      &      2     &   3    &  \\
    $\sigma$ (km s$^{-1}$) &  $ 50 \pm 10$ & $55 \pm 15$  &  $55 \pm 20$  &  & $50 \pm 5$ & $105 \pm 30$  &  $135 \pm 50$ & \\
    $v_{\rm out}$ (km s$^{-1}$) & $-270 \pm 40$ & $-545 \pm 55$  &  $-820 \pm 80$ &  & $-70 \pm 40^a$ & $-380 \pm 60^a$ & $-880 \pm 90^a$ & \\
    \hline
    Ion  &  $\log N_{\rm ion}$  & $\log N_{\rm ion}$  & $\log N_{\rm ion}$  &   $\log \xi$  & $\log N_{\rm ion}$  &  $\log N_{\rm ion}$  & $\log N_{\rm ion}$  &   $\log \xi$ \\
        &  (cm$^{-2}$) & (cm$^{-2}$) & (cm$^{-2}$) & (erg cm s$^{-1}$) &  (cm$^{-2}$) & (cm$^{-2}$) & (cm$^{-2}$) & (erg cm s$^{-1}$) \\
    (1) &    (2)  &  (3)  & (4)  &  (5)   &  (6)  &  (7)  &  (8)  &  (9) \\
    \hline
    \CV{}  & $16.9 \pm 0.4$ & $16.1 \pm 0.7$ & $<15.4$ & $0.30$ & $<16.7$ & $<15.7$ & $16.0 \pm 0.5$ & $-0.15$ \\
    \CVI{} & $<15.6$ & $16.9 \pm 0.4$ & $<16.1$ & $1.20$ & $<15.0$ & $17.1 \pm 0.2$ & $<15.4$ & $1.00$ \\
    \NV{}  & $<15.1$  & $<14.9$ & $<14.9$ & $0.15$ & $<15.7$ & $15.8 \pm 0.7$ & $<15.4$ & $0.00$ \\
    \NVI{} & $16.6 \pm 0.3$  & $<16.1$ & $<15.4$ & $0.80$ & $<16.7$ & $<17.2$ & $15.5 \pm 0.4$ & $0.70$ \\
    \NVII{} & $16.3 \pm 0.6$ & $<15.6$ & $16.2 \pm 0.5$ & $1.50$ & $16.8 \pm 0.5$ & $<15.5$ & $15.8 \pm 0.4$ & $1.35$ \\
    \OV{}  & $<16.0$  & $16.4 \pm 0.4$ & $<16.7$ & $0.20$ & $15.9 \pm 0.5$ & $15.9 \pm 0.6$ & $<15.3$ & $0.00$ \\
    \OVI{} & $15.9 \pm 0.5$  & $15.9 \pm 0.6$ & $<15.9$ & $0.50$ & $16.2 \pm 0.3$ & $<16.5$ & $<15.4$ & $0.40$ \\
    \OVII{} & $17.6 \pm 0.1$ & $15.7 \pm 0.6$ & $<16.0$ & $1.10$ & $17.8 \pm 0.1$ & $15.8 \pm 0.6$ & $15.8 \pm 0.4$ & $0.95$ \\
    \OVIII{} & $18.0 \pm 0.1$ & $<16.1$ & $<16.1$ & $1.75$ & $17.8 \pm 0.1$ & $<15.5$ & $<15.4$ & $1.60$ \\
    \NeIX{} & $<16.5$ & $17.1 \pm 0.3$ & $<16.0$ & $1.60$ & $17.0 \pm 0.7$ & $17.5 \pm 0.6$ & $<16.0$ & $1.50$ \\
    \MgX{} & $<15.7$ & $<16.4$ & $<16.7$ & $1.70$ & $<16.5$ & $17.0 \pm 0.4$ & $<16.2$ & $1.50$ \\
    \SXII{} & $16.4 \pm 0.4$ & $<15.7$ & $15.8 \pm 0.5$ & $2.10$ & $15.8 \pm 0.5$ & $<15.5$ & $<15.3$ & $1.90$ \\
    \FeIX{}  & $15.7 \pm 0.3$ & $<15.7$ & $15.5 \pm 0.5$ & $1.40$ & $<15.5$ & $<15.8$ & $<15.5$ & $1.25$ \\
    \FeXVI{} & $<15.6$ & $<15.1$ & $16.4 \pm 0.2$ & $2.10$ & $<14.8$ & $<14.9$ & $<14.8$ & $1.90$ \\
    \FeXVII{} & $<15.9$ & $<16.3$ & $16.3 \pm 0.3$ & $2.20$ & $15.3 \pm 0.8$ & $16.0 \pm 0.5$ & $<14.8$ & $2.00$ \\
    \FeXVIII{} & $<16.1$ & $<15.6$ & $15.8 \pm 0.3$ & $2.30$ & $16.2 \pm 0.2$ & $<16.2$ & $<15.2$ & $2.20$ \\
    \FeXIX{} & $16.0 \pm 0.5$ & $<16.4$ & $15.9 \pm 0.4$ & $2.50$ & $16.3 \pm 0.3$ & $<16.3$ & $<16.1$ & $2.45$ \\
    \FeXX{} & $<15.3$ & $<15.6$ & $16.5 \pm 0.2$ & $2.70$ & $<15.9$ & $<16.1$ & $16.2 \pm 0.2$ & $2.70$ \\
    \FeXXI{} & $<15.7$ & $16.7 \pm 0.5$ & $<15.8$ & $2.90$ & $<16.7$ & $<16.9$ & $16.8 \pm 0.3$ & $2.90$ \\
    \FeXXII{} & $<16.6$ & $<16.4$ & $16.0 \pm 0.6$ & $3.05$ & $<15.7$ & $<15.7$ & $<15.7$ & $3.05$ \\
    \hline
  \end{tabular}
  \medskip
  {\bf Note:} $^a$To compare with the LETGS wavelength scale, $110 \pm 55$~km s$^{-1}$ must be substracted to these values (see text).
  \end{minipage}
\end{table*}

\subsubsection{Absorption lines: photoionisation modeling}
\label{xrayxabs}

The {\it slab} model suggests that the absorption system in NGC 4593 may consist of different components in different kinematic regimes and ionisation states. Since we did not detect significant variability in the properties of the absorbers in Sect.~\ref{xrayslab}, we modeled the absoption spectrum simultaneously for the RGS and LETGS spectra with the {\it xabs} model, which calculates the transmission of a slab of material where all ionic column densities are linked through a photoionisation balance model. This was calculated using {\it CLOUDY} (\citealt{Fer98}) version C10. The fitting package {\it SPEX} allows us to place the spectra in different sectors so that the {\it xabs} free parameters can be coupled, while the continuum and emission line parameters are independently fitted. The photoionsation balances provided to {\it xabs} calculated from the SED at the epochs of the \XMM{}~and \Chan{}~observations are also assigned to each sector independently.

The free parameters in the {\it xabs} fit are the ionisation parameter $\xi$, the hydrogen column density $N_{\rm H}$, the velocity width $\sigma$, and the outflow velocity $v_{\rm out}$. The abundances were set to the proto-Solar values of \citet{LP09}. The ionisation parameter $\xi$ is a measure of the ionisation state of the gas, and it is defined as

\begin{equation}
  \label{xieq}
  \xi=L/nR^2,
\end{equation}

\noindent where $L$ is the ionising luminosity in the $1-1000$~Ryd range, $n$ is the hydrogen density of the gas, and $R$ is the distance from the ionising source.

\begin{figure}
  \includegraphics[width=6.5cm,angle=-90]{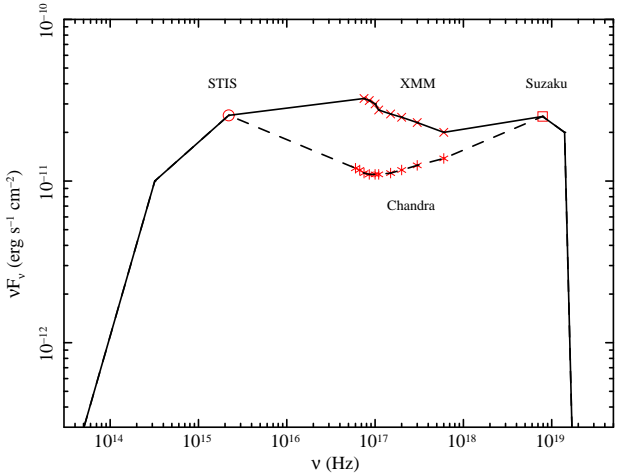}
  \caption{Spectral energy distribution of NGC 4593.}
  \label{sed}
\end{figure}

The spectral energy distribution (SED) provided to {\it CLOUDY} as an input was built using the X-ray and UV points from the \XMM{}, \Chan{} and STIS observations discussed in this paper. At higher frequencies we use an observation of NGC 4593 by {\it Suzaku} in 2007 (\citealt{MR09}). At lower energies, the shape of the SED resembles the one assumed by {\it CLOUDY} for a standard ionisation continuum. We note that only the \STIS{}~and \XMM{}~observations were taken simultaneously. However, the SED with \STIS{}~and \Chan{}~data may still provide reliable results since the bulk of the ionising continuum is in the X-ray domain and that the UV continuum variability is typically less than 60\%, although changes of up to a factor of 4 can occasionally be seen, according to {\it IUE} monitoring for almost a decade\footnote{\tt http://www.chara.gsu.edu/PEGA/IUE/ObjectTbl.cgi} (\citealt{Dunn06}). The adopted SED is shown in Fig.~\ref{sed}.

\begin{figure*}
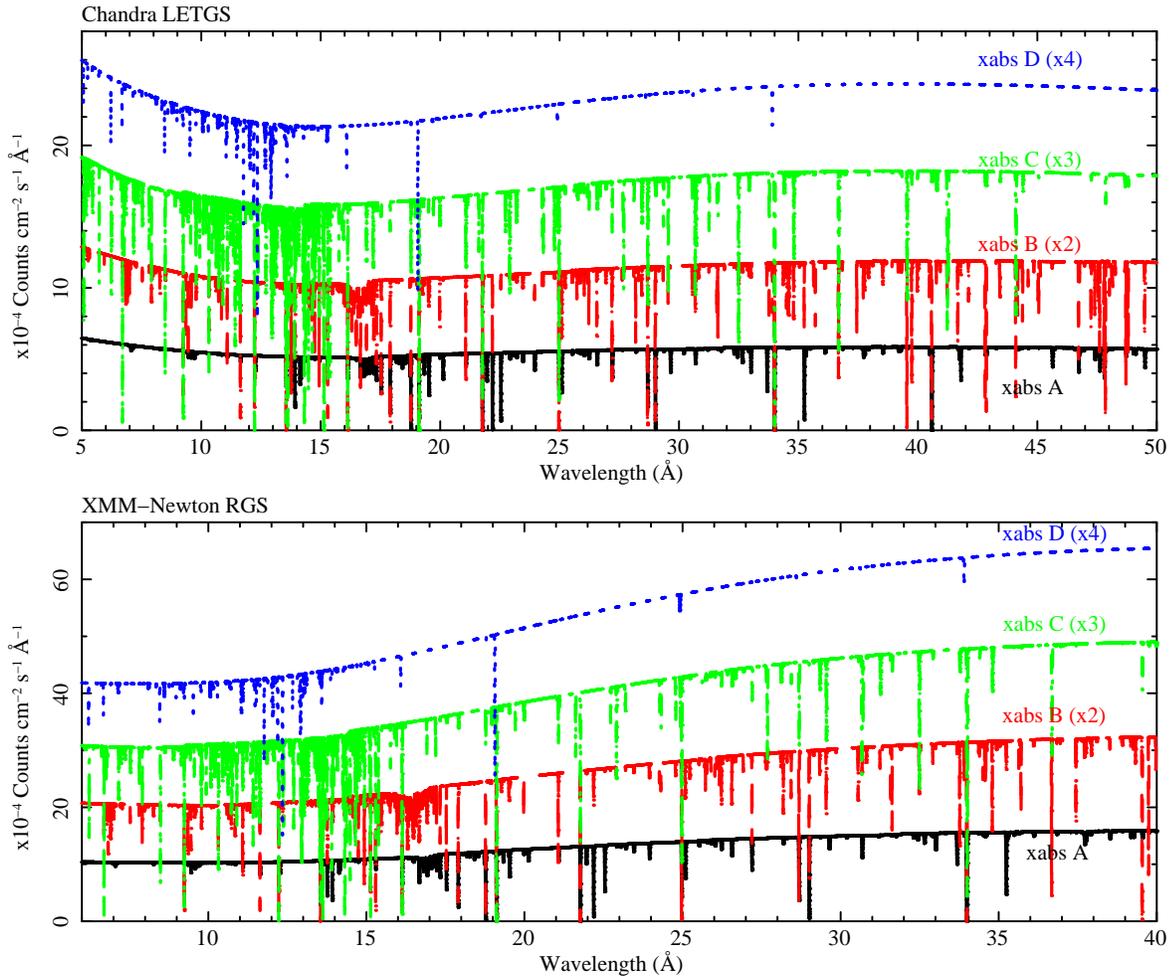

  \hbox{
    \includegraphics[width=6.5cm,angle=-90]{letgs_xabs.ps}
  }
  \hbox{
    \includegraphics[width=6.5cm,angle=-90]{rgs_xabs.ps}
  }
  \caption{Model plot showing the contribution of each of the {\it xabs} WA components to the \Chan{}~LETGS (upper panel) and the \XMM{}~RGS (lower panel) spectra. In both cases the solid line represents WA component A, while the dashed, dot-dashed, and dotted lines represent WA components B, C, and D, respectively. For clarity, components B, C, and D have been shifted in flux by a factor 2, 3, and 4, respectively. The wavelength scales are in the observer frame.}
  \label{xabsmodel}
\end{figure*}

As in the case of {\it slab} we initially fitted the spectra using a single {\it xabs} component in addition to the continuum and emission line models. This hugely improved the overall fit in $\Delta C/\nu = 292/4$, albeit several troughs were not correctly accounted for. We therefore added a second {\it xabs} component that further improved the fit by $\Delta C/\nu = 201/4$, but it still left some residuals in absorption. Since the {\it slab} modeling suggested the presence of three kinematic components we added a third {\it xabs} component to the fit, which was improved by $\Delta C/\nu = 37/4$, well above the $4\sigma$ level. The WA structure at this stage was characterized by two low ionisation components with $\log \xi \sim 1.0$ and $\log \xi \sim 1.7$, respectively, and a higher ionisation component with $\log \xi \sim 2.7$. The lower ionisation components are almost one order of magnitude shallower than the high-$\xi$ one ($N_{\rm H}$ of a few times $10^{20}$~cm$^{-2}$ against $N_{\rm H} \sim 3 \times 10^{21}$~cm$^{-2}$). Interestingly, the fit further improves above the $3\sigma$ level ($\Delta C/\nu = 28/4$) after adding a fourth {\it xabs} component, which resulted in the high-$\xi$ component being split in two components with $\log \xi \sim 2.4$ and $\log \xi \sim 3.0$, with $N_{\rm H} \sim 2.5 \times 10^{21}$~cm$^{-2}$ and $N_{\rm H} \sim 5 \times 10^{20}$~cm$^{-2}$, respectively. The kinematic structure shows, as suggested by the {\it slab} modeling, increasing outflow velocities as the ionisation parameter increases. The addition of more {\it xabs} components no longer improved the fit, so a 4-component WA was adopted as the final best-fit solution with ${\rm C^2}/\nu = 2878/2591$. The best-fit values are reported in Table~\ref{WAbest}. A model plot of the different {\it xabs} components in the LETGS and RGS spectra is shown in Fig.~\ref{xabsmodel}.

The results of these fits reveal that the WA in NGC 4593 is made of 4 components in distinct ionisation states, namely $\log \xi_{\rm A} = 1.0 \pm 0.1$, $\log \xi_{\rm B} = 1.7 \pm 0.1$, $\log \xi_{\rm C} = 2.4 \pm 0.1$, and $\log \xi_{\rm D} = 3.0 \pm 0.2$. In what follows these components will be labeled as $A$, $B$, $C$, and $D$, where $A$ is the lowest and $D$ is the highest ionisation gas. We observe a mild but significant correlation between the ionisation degree of the absorbers and their column density; the more ionised is the gas, the larger is the column density except for component $D$ which has a somewhat lower $N_{\rm H}$ albeit with large uncertainties. Such a behaviour is not uncommon among the WA detected in other Seyfert galaxies. A similar correlation seems to exist between the ionisation parameter and the outflow velocity. The kinematic structure of the outflow suggests the presence of three kinematic components: the absorbers $A$ and $B$ are outflowing at $v_{\rm out} \sim -350$~km s$^{-1}$, while component $C$ is outflowing at $v_{\rm out} \sim -450$~km s$^{-1}$, and the highest ionisation one is much faster with $v_{\rm out} \sim -1100$~km s$^{-1}$. This behaviour may provide a hint on the location where these outflows were formed, as it is typically believed that a larger outflow velocity might indicate that the wind was launched closer to the central ionising source. This will be further discussed in Sect.~\ref{discussion}.

\begin{table}
  \caption{Warm absorber best-fit parameters. Column (1): WA component; column (2): ionisation parameter; column (3): hydrogen column density; column (4): velocity width; column (5): outflow velocity.}
  \label{WAbest}
  \begin{tabular}{@{}lcccc}
    \hline
    Component  &  $\log \xi$  &  $N_{\rm H}$  &  $\sigma$  &  $v_{\rm out}$ \\
               & (erg cm s$^{-1}$) &  (cm$^{-2}$) & (km s$^{-1}$) &  (km s$^{-1}$) \\
    (1)        &  (2)             & (3)          & (4)          & (5) \\
    \hline
    $A$        &  $1.0 \pm 0.1$  &  $3.0 \pm 0.6 \times 10^{20}$ & $80 \pm 15$ & $-370 \pm 35$ \\
    $B$        &  $1.7 \pm 0.1$  &  $6.5 \pm 0.9 \times 10^{20}$ & $40 \pm 20$ & $-335 \pm 55$ \\
    $C$        &  $2.4 \pm 0.1$  &  $2.4 \pm 0.1 \times 10^{21}$ & $75 \pm 20$ & $-460 \pm 55$ \\
    $D$        &  $3.0 \pm 0.2$  &  $5.5_{-2.5}^{+8.5} \times 10^{20}$ & $75 \pm 50$ & $-1150 \pm 200$ \\
    \hline
  \end{tabular}
\end{table}

\subsection{UV: HST/STIS}
\label{stis}

\begin{figure*}
  \includegraphics[width=6.5cm,angle=-90]{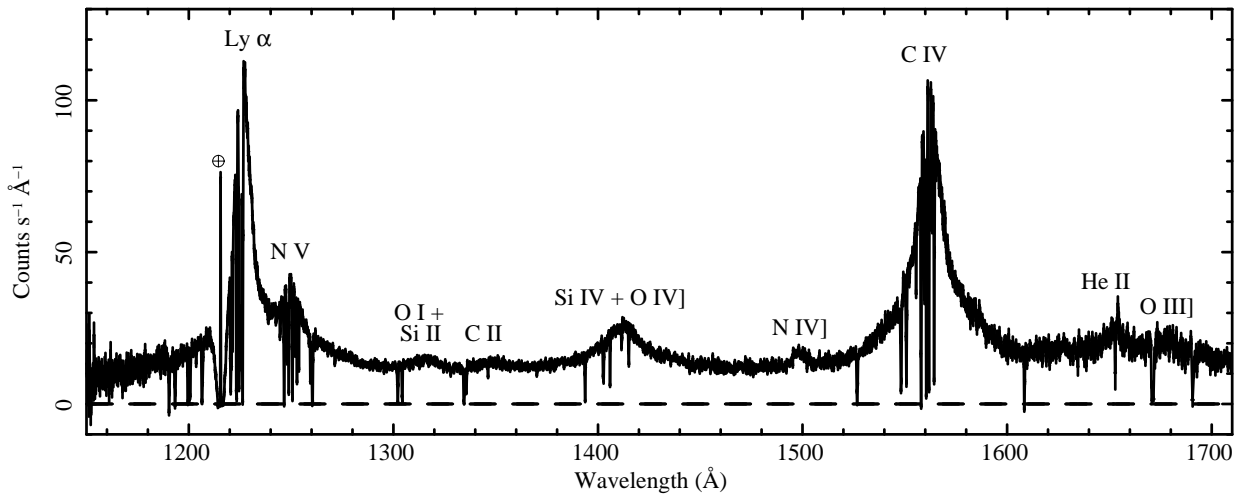}
  \caption{HST/STIS spectrum of NGC 4593. Some of the most relevant emission features are labeled. An atmospheric line is located to the left of the \Lya{} complex. The wavelength scale is in the observer frame.}
  \label{stisspec}
\end{figure*}

\begin{figure*}
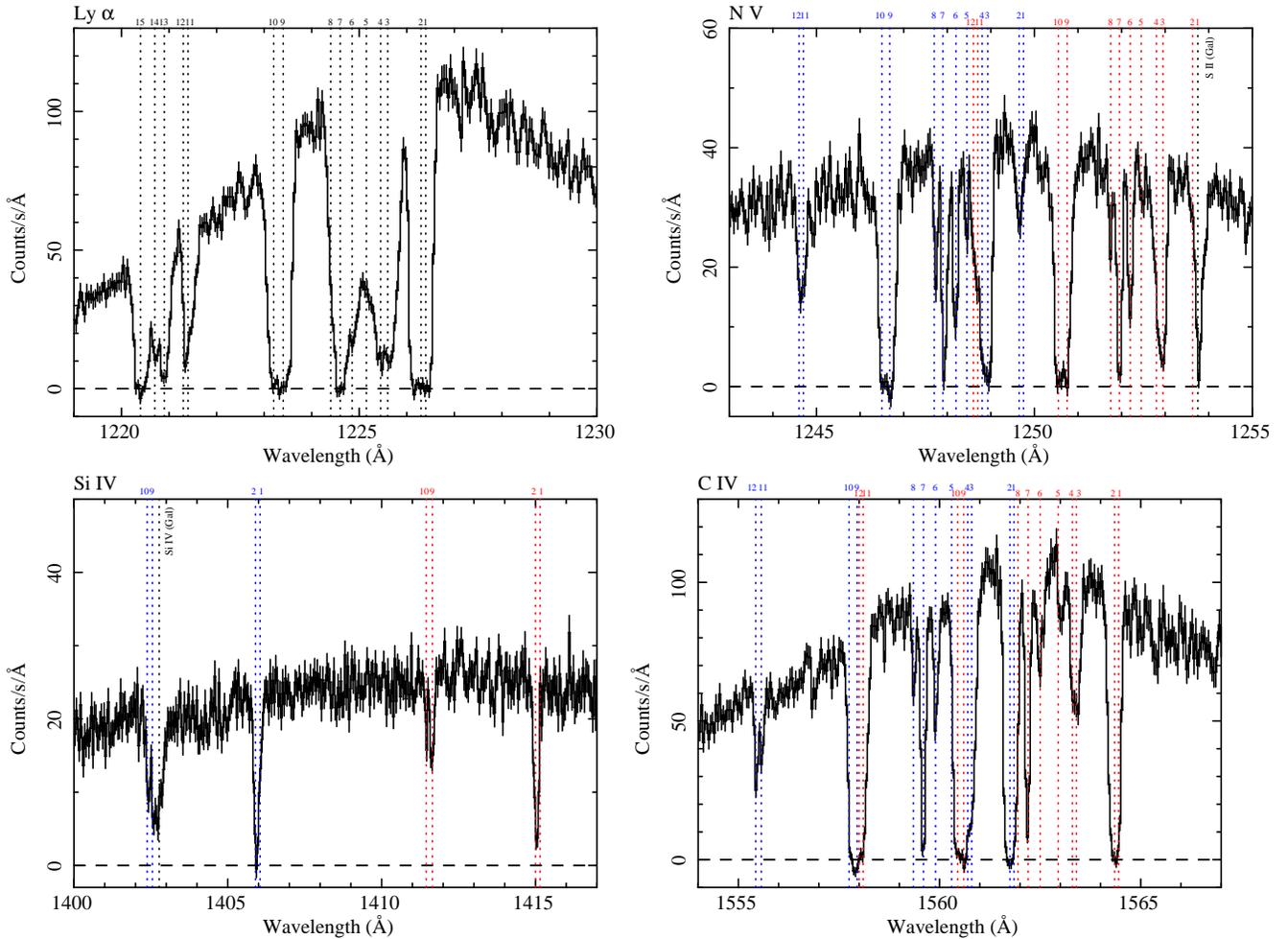

  \hbox{
    \includegraphics[width=6.5cm,angle=-90]{spec_Lya.ps}
    \includegraphics[width=6.5cm,angle=-90]{spec_NV.ps}
  }
  \hbox{
    \includegraphics[width=6.5cm,angle=-90]{spec_SiIV.ps}
    \includegraphics[width=6.5cm,angle=-90]{spec_CIV.ps}
  }
  \caption{Detail of the \STIS{} spectrum in the \Lya{} (upper panel left), \NV{} (upper panel right), \SiIV{} (lower panel left), and \CIV{} (lower panel right) regions. The different kinematic components found in the absorber are marked. The wavelength scales are in the observer frame.}
  \label{speclya}
\end{figure*}

The UV spectrum of NGC 4593 obtained by HST/STIS shows a wealth of emission and absorption lines in the $1150-1710$~\AA~waveband. The most prominent emission features correspond to \Lya{} and \CIV{}, but others belonging to \NV{}, \SiII{}, \SiIV{}, and \NIV{} can be clearly seen. A complex absorption spectrum is superimposed to these emission lines, with several troughs produced by intervening gas at the redshift of the source while others belong to the local interstellar medium (ISM) and the intergalactic medium (IGM) between our Galaxy and NGC 4593 (see Fig.~\ref{stisspec}).

A detailed analysis of the UV continuum and physical properties of the ISM and IGM intervening gas is out of the scope of this paper, so we will focus only on the properties of the absorbing gas intrinsic to NGC 4593. The regions of interest are those around the position of \Lya{} and the \NV{}, \SiIV{}, and \CIV{} doublets. For each of them the continuum was modeled locally with a power law and with Gaussians for the emission lines. For the characterization of the absorption troughs we first used a kinematic approach using {\it slab}, which is model-independent, in a similar manner as described in Sect.~\ref{xrayslab}. We used one {\it slab} component per absorption trough and we left the velocity shift (i.e. outflow velocity $v_{\rm out}$), the {\it rms} velocity broadening $\sigma$, the ionic column density $N_{\rm ion}$ of the species under consideration, and the covering factor of the absorber $f_{\rm C}$ as free parameters in the fit.

\begin{figure}
  \includegraphics[width=8cm,angle=0]{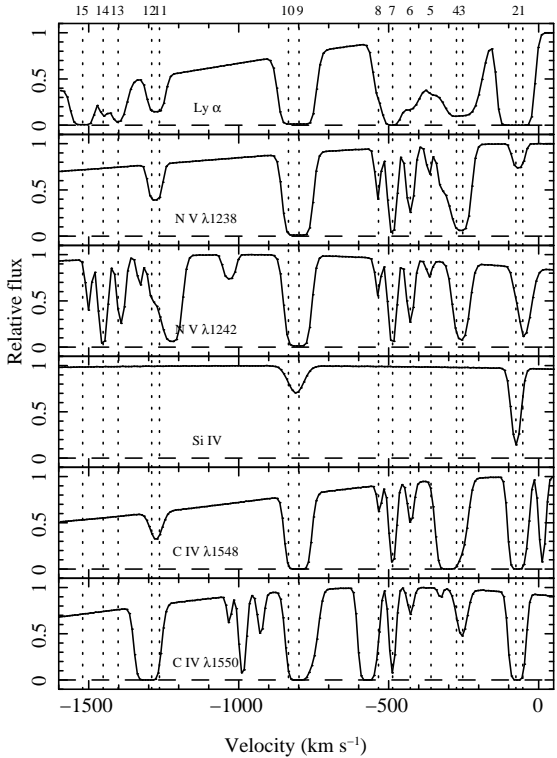}
  \caption{Intrinsic absorption features in the STIS spectrum of NGC 4593. Normalized relative fluxes are plot as a function of velocity relative to the systemic redshift of $z = 0.009$. The panels show in descending order the troughs in \Lya{}, the blue and red sides of the \NV{} doublet, \SiIV{}, and the blue and red sides of the \CIV{} doublet. Vertical dotted lines mark the different kinematic components.}
  \label{stisuv}
\end{figure}

A total of 15 different kinematic components can be identified in the STIS spectrum of NGC 4593. Twelve of these components are found in common in \CIV{}, \NV{}, and \Lya{}, and only four of them are seen in \SiIV{}. In addition to them, three faster components with $v_{\rm out}$ ranging between $-1400$ and $-1520$~km s$^{-1}$ are identified only in the \Lya{} region. In what follows kinematic components are numbered by increasing outflow velocity, being component 1 the slowest, and component 15 the fastest. A detailed view of the STIS spectrum in the \Lya{}, \NV{}, \SiIV{}, and \CIV{} regions is shown in Fig.~\ref{speclya}, where all the detected kinematic components are labeled together with two foreground absorption features of S\,{\sc ii} $\lambda 1253.81$ and \SiIV{} $\lambda 1402.77$. The different kinematic components are also shown in velocity space in Fig.~\ref{stisuv}.

From the best-fit results (see Table~\ref{uvslab}) we can see that the bulk of the ionised gas in the UV is carried by components 1 and 9, as the ionic column densities of \CIV{} reach a maximum in these components, and similarly for \NV{} in components 3 and 7. We note, however, that some of these components are heavily saturated, namely component 1 in \CIV{} and component 9 in both \NV{} and \CIV{}, and hence only lower limits on their ionic column densities can be reliably determined. Likewise, the column density of the \Lya{} absorber is maximal for components 3, 7, and 10, and their $N_{\rm Ly\,\alpha}$ values should be considered as lower limits due to the large degree of saturation. The errors on $v_{\rm out}$ and $\sigma$ quoted in Table~\ref{uvslab} account for both the statistical and systematic uncertainties\footnote{\tt http://www.stsci.edu/hst/stis/performance/spectral\_resolution}.

After determining the kinematic structure of the UV absorber, we used a more physical model to characterize its ionisation properties. For this purpose we locally fitted the spectrum using {\it xabs} components using the best-fit values of $v_{\rm out}$, $\sigma$, and $f_{\rm C}$ obtained with {\it slab} as a starting point. We found that the fastest absorption troughs 13 to 15, seen only in the \Lya{} region, are caused by a shallow ($N_{\rm H} \lesssim 10^{17}$~cm$^{-2}$) neutral absorber ($\log \xi \sim -0.5$ on average) along our line of sight. The different kinematic components seen in \NV{}, \SiIV{} and \CIV{} are associated to mildly ionised gas ($\log \xi \sim 0.6$ for components 8 to 12, and $\log \xi \sim 0.2$ for component 1), with the exception of components 3 to 7  which might originate in a somewhat more ionised environment ($\log \xi \sim 1.6$). The best-fit results are shown in Table~\ref{uvxabs}. In Sect.~\ref{discussion} we discuss these results in more depth, and the possibility that the UV absorber and the low ionisation component $A$ seen in X-rays could originate in the same enviroment as they seem to share several physical properties.

\begin{table*}
  \caption{Intrinsic absorption features in the STIS spectrum of NGC 4593. Column (1): UV kinematic component; column (2): outflow velocity; column (3): velocity width; column (4): \Lya{} column density; column (5): \NV{} column density; column (6): \SiIV{} column density; column (7): \CIV{} column density; column (8): covering factor.}
  \label{uvslab}
  \begin{tabular}{@{}lccccccc}
    \hline
    Component  &  $v_{\rm out}$  &  $\sigma$  &  $\log N_{\rm Ly\,\alpha}$  & $\log N_{\rm N\,V}$  & $\log N_{\rm Si\,IV}$  & $\log N_{\rm C\,IV}$  &  $f_{\rm C}$ \\
               &   (km s$^{-1}$) & (km s$^{-1}$) & (cm$^{-2}$)  & (cm$^{-2}$)  & (cm$^{-2}$)  & (cm$^{-2}$)  &    \\
    (1)    &  (2)              & (3)        & (4)        & (5) &  (6)  &  (7) &  (8)  \\
    \hline
    1 &  $-61 \pm 3$ & $11 \pm 3$ & $>14.6$ & $13.05 \pm 0.32$ & $13.48 \pm 0.16$ & $>15.5$ & $>0.997$ \\
    2 &  $-76 \pm 3$ & $16 \pm 3$ & $>15.9$ & $13.10 \pm 0.33$ & $13.54 \pm 0.11$ & $12.87 \pm 0.49$ & $>0.993$ \\
    3 &  $-253 \pm 3$ & $13 \pm 4$ & $>16.5$ & $14.87 \pm 0.07$ & $\dots$ & $13.88 \pm 0.04$ & $0.853 \pm 0.008$ \\
    4 &  $-274 \pm 5$ & $23 \pm 5$ & $13.36 \pm 0.10$ & $14.16 \pm 0.05$ & $\dots$ & $13.42 \pm 0.12$ & $>0.935$ \\
    5 &  $-359 \pm 4$ & $12 \pm 4$ & $16.43 \pm 0.25$ & $13.57 \pm 0.09$ & $\dots$ & $<12.0$ & $0.578 \pm 0.014$ \\
    6 &  $-428 \pm 3$ & $9 \pm 3$ & $16.21 \pm 0.25$ & $14.53 \pm 0.08$ & $\dots$ & $13.52 \pm 0.05$ & $0.778 \pm 0.013$ \\
    7 &  $-487 \pm 3$ & $10 \pm 3$ & $>16.3$ & $14.78 \pm 0.06$ & $\dots$ & $14.25 \pm 0.03$ & $>0.998$ \\
    8 &  $-534 \pm 4$ & $10 \pm 4$ & $13.37 \pm 0.05$ & $13.62 \pm 0.05$ & $\dots$ & $13.10 \pm 0.06$ & $>0.953$ \\
    9 &  $-799 \pm 3$ & $17 \pm 3$ & $15.85 \pm 0.06$ & $>15.5$ & $13.30 \pm 0.04$ & $>14.9$ & $>0.995$ \\
    10 & $-834 \pm 3$ & $11 \pm 3$ & $>16.0$ & $14.68 \pm 0.12$ & $12.90 \pm 0.09$ & $14.15 \pm 0.06$ & $0.868 \pm 0.023$ \\
    11 & $-1264 \pm 4$ & $11 \pm 4$ & $13.79 \pm 0.06$ & $13.96 \pm 0.09$ & $\dots$ & $13.60 \pm 0.08$ & $0.524 \pm 0.023$ \\
    12 & $-1290 \pm 4$ & $10 \pm 4$ & $13.60 \pm 0.06$ & $13.76 \pm 0.10$ & $\dots$ & $13.52 \pm 0.11$ & $0.882 \pm 0.033$ \\
    13 & $-1402 \pm 4$ & $18 \pm 4$ & $14.00 \pm 0.04$ & $\dots$ & $\dots$ & $\dots$ & $>0.979$ \\
    14 & $-1452 \pm 5$ & $11 \pm 5$ & $13.56 \pm 0.05$ & $\dots$ & $\dots$ & $\dots$ & $>0.918$ \\
    15 & $-1520 \pm 4$ & $15 \pm 3$ & $>15.0$ & $\dots$ & $\dots$ & $\dots$ & $>0.990$ \\
    \hline
  \end{tabular}
\end{table*}

\begin{table*}
  \caption{Best-fit photoionisation model for the UV absorbers. Column (1): UV kinematic component; column (2): outflow velocity; column (3): velocity width; column (4): ionisation parameter; column (5): hydrogen column density; column (6): covering factor.}
  \label{uvxabs}
  \begin{tabular}{@{}lccccc}
    \hline
    Component & $v_{\rm out}$  &  $\sigma$  &  $\log \xi$ &  $\log N_{\rm H}$  &  $f_{\rm C}$ \\
               & (km s$^{-1}$) & (km s$^{-1}$) & (erg cm s$^{-1}$) & (cm$^{-2}$)  &    \\
    (1)       & (2)           & (3)           & (4)              & (5)          & (6) \\
    \hline
     1 & $-57 \pm 3$  & $9 \pm 3$ & $0.23 \pm 0.03$  & $18.36 \pm 0.02$   & $>0.995$ \\
     2 & $-81 \pm 3$  & $11 \pm 3$ & $-0.18 \pm 0.02$  & $18.73 \pm 0.03$   & $>0.998$ \\
     3 & $-254 \pm 4$  & $13 \pm 3$ & $1.68 \pm 0.01$  & $20.32 \pm 0.02$   & $0.806 \pm 0.010$ \\
     4 & $-279 \pm 5$  & $18 \pm 4$ & $1.74 \pm 0.02$  & $20.17 \pm 0.05$   & $0.974 \pm 0.021$ \\
     5 & $-358 \pm 6$  & $9 \pm 5$ & $1.51 \pm 0.03$  & $17.65 \pm 0.05$   & $>0.911$ \\
     6 & $-428 \pm 4$  & $10 \pm 4$ & $1.57 \pm 0.01$  & $19.56 \pm 0.03$   & $0.750 \pm 0.012$ \\
     7 & $-487 \pm 3$  & $9 \pm 3$ & $1.10 \pm 0.02$  & $19.32 \pm 0.03$   & $>0.987$ \\
     8 & $-532 \pm 4$  & $8 \pm 4$ & $0.81 \pm 0.08$  & $18.55 \pm 0.09$   & $0.352 \pm 0.018$ \\
     9 & $-796 \pm 3$  & $12 \pm 3$ & $0.74 \pm 0.01$  & $19.94 \pm 0.03$   & $0.980 \pm 0.006$ \\
     10 & $-832 \pm 3$  & $10 \pm 3$ & $0.62 \pm 0.02$  & $19.24 \pm 0.05$   & $0.886 \pm 0.014$ \\
     11 & $-1262 \pm 5$  & $11 \pm 4$ & $0.80 \pm 0.11$  & $18.81 \pm 0.11$   & $0.305 \pm 0.021$ \\
     12 & $-1287 \pm 4$  & $8 \pm 4$ & $0.89 \pm 0.05$  & $18.23 \pm 0.07$   & $0.594 \pm 0.026$ \\
     13 & $-1402 \pm 4$  & $13 \pm 4$ & $0.22 \pm 0.03$  & $16.89 \pm 0.03$   & $>0.982$ \\
     14 & $-1452 \pm 4$  & $12 \pm 5$ & $-0.72 \pm 0.04$  & $15.43 \pm 0.04$   & $>0.952$ \\
     15 & $-1520 \pm 4$  & $10 \pm 4$ & $-0.27 \pm 0.08$  & $17.32 \pm 0.06$   & $>0.991$ \\
    \hline
  \end{tabular}
\end{table*}

\section{Discussion}
\label{discussion}

\subsection{Comparison with previous X-ray observations}
\label{comparison}

We first compare these results with those of other previous X-ray observations of NGC 4593. The source was observed by \XMM{} in July 2000 and by \Chan{} LETGS in February 2001. The analysis of these observations are reported in \citet{Steen03}. Because of the high radiation background, the \XMM{} data were severely affected and the RGS spectrum was rather noisy. However, \citet{Steen03} were able to derive the ionic column densities of several species present in the spectrum albeit with large uncertainties. Their results are fully in agreement with ours (reported in Table~\ref{slab}) within the error bars. Their analysis of the LETGS data revealed the presence of highly ionised gas ($\log \xi = 2.61 \pm 0.09$) outflowing at $-400 \pm 120$~km s$^{-1}$ with a column density of $N_{\rm H} = 1.6 \pm 0.4 \times 10^{21}$~cm$^{-2}$. Their results are hence consistent with what we found for component $C$ in the outflow at the $1\sigma$ confidence level, although we detect a marginally higher column density for this component. As in our observations, they required an extra component to account for weak absorption lines in the oxygen region resulting in a shallow ($N_{\rm H} = 6 \pm 3 \times 10^{19}$~cm$^{-2}$) lower ionisation gas ($\log \xi = 0.5 \pm 0.3$) outflowing at $-380 \pm 140$~km s$^{-1}$. Again, these results are in agreement with those of our lowest ionisation component $A$. The outflow velocities of the components detected by \citet{Steen03} are consistent with each other within $1\sigma$, similarly to what we find for components $A$ to $C$. Their available statistics made them unable to discriminate whether these outflows belonged to a single kinematic component.

The ionised gas in NGC 4593 was also observed with the \Chan{} High Energy Transmission Gratings (HETGS) in June 2001 (\citealt{McK03}). Since they cover a higher energy range, a single-zone photoionised absorber suffices to adequately describe the soft X-ray spectrum of NGC 4593. The reported parameters of this gas reported in \citet{McK03} are $\log \xi = 2.52 \pm 0.04$ and $N_{\rm H} = 5.4 \pm 0.9 \times 10^{21}$~cm$^{-2}$ (for consistency, we have converted their quoted 90\% confidence errors to 1$\sigma$). Although the value of the ionisation parameter $\xi$ is in agreement with the one we derive for component $C$ in our dataset, they measure a much slower outflow velocity of $-140 \pm 35$~km~s$^{-1}$ and a column density $N_{\rm H}$ higher by a factor of 2.

NGC 4593 was observed again by \XMM{} in June 2002. The results of this observation are reported in \citet{Bre07}. They did not use the RGS spectrum (analysed in this paper) and they relied only on the analysis of the EPIC-pn spectrum. Even though the lower spectral resolution of EPIC-pn does not allow to extract information on individual lines, \citet{Bre07} were able to derive the ionisation parameter and column densities of the intervening absorber. Their results depended on the model used to parametrize the soft excess: single temperature bremsstrahlung emission or thermal Comptonization by plasma at a temperature between that of the disc and the hard X-ray corona. In either case they require two ionised absorbers at the redshift of the source to reproduce the observed spectrum. The former yielded two components with $\log \xi = 2.54 \pm 0.02$ and $N_{\rm H} = 1.64 \pm 0.05 \times 10^{23}$~cm$^{-2}$, and $\log \xi = 0.57 \pm 0.09$ and $N_{\rm H} = 2.5^{+0.2}_{-0.4} \times 10^{21}$~cm$^{-2}$, respectively. The latter provides a best fit with $\log \xi = 2.75^{+0.07}_{-0.19}$ and $N_{\rm H} = 9.3^{+0.9}_{-5.6} \times 10^{22}$~cm$^{-2}$, and $\log \xi = 1.70^{+0.18}_{-0.36}$ and $N_{\rm H} = 1.13^{+0.73}_{-0.57} \times 10^{22}$~cm$^{-2}$, respectively (as above, their quoted 90\% errors have been converted into 1$\sigma$ errors for consistency). While these results indicate a somewhat deeper absorber ($N_{\rm H}$ higher than ours by a factor of $\sim$5 in the first case), their ionisation parameters are generally consistent with those of our components. Therefore, it is clear from this analysis and the previous ones that the WA in NGC 4593 can be successfully described with at least two distinct components, although a more complex structure cannot be ruled out. The kinematic distribution of these components remain, however, somewhat unclear.

\subsection{Emission line diagnostics}
\label{emission}

Some emission features of oxygen, nitrogen, and carbon are seen in the soft X-ray spectra of NGC 4593 in the \XMM{} and \Chan{} observations, as reported in Sect.~\ref{xraycont}. Lines arising in He-like species like \NeIX{}, \OVII{}, and \NVI{} are of particular interest. In the case of \OVII{} and \NVI{} only the forbidden ($f$) and intercombination ($i$) lines of the triplets are significantly detected, while for \NeIX{} the only significant feature is the $f$ line. The lack of a significant detection of resonance ($r$) lines together with relatively strong $f$ lines may indicate that the dominant ionising process photoionisation (\citealt{PD00}). We note, however, that we are unable to significantly detect any radiative recombination continua (RRC) features in the spectrum, which are typical signatures in photoionised plasmas.

Emission line diagnostics can be used to estimate some of the physical properties of the line emitting gas. The $R = f/i$ ratio is often used as an estimator of the electron density of the gas. Using the \OVII{} triplet, ratios $R(n_e) = 1.8 \pm 0.9$ and $R(n_e) = 1.9 \pm 1.2$ are measured in the \XMM{} and \Chan{} observations, respectively. This result, taking into account the large uncertainties, points to emission in a gas with density between $n_e \sim 10^{10}$~and $n_e \sim 10^{11}$~cm$^{-3}$, according to \citet{PD00}. The \NVI{} emission lines in the LETGS spectrum, detected only at the $\sim 2\sigma$ level, yields $R(n_e) = 1.4 \pm 1.6$, formally an upper limit. The RGS spectrum, however, shows a somewhat contradictory result as the $i$ line is much stronger than the $f$ line and $R(n_e) = 0.4 \pm 0.2$, thus implying that this emission feature arises in a higher density plasma $n_e \sim 10^{11}$~cm$^{-3}$. In principle the ratio $G = (f+i)/r$ can also be used to estimate the electron temperature of the gas. However due to the non-detection of the $r$ line in neither of the observations and triplets, it is not feasible to derive an accurate $G$ ratio for NGC 4593.

The line fluxes of the \OVII{} and \NVI{} $f$ and $i$ lines decreased between the \XMM{}~and \Chan{} observations (see Tables~\ref{cont}~and~\ref{rgscont}). If these changes are caused in response to the variations in the flux of the source, an upper limit to the distance of the emitter can be immediately estimated by means of the light distance traveled. Given the time span between both observations (3147 days at the rest frame of NGC 4593) we derive a distance of $< 2.5$~pc.

Since the observations were taken more than 8 years apart, this somewhat loose upper limit cannot be further constrained with the available data. The observations taken with \Chan{} in 2001 and that of \XMM{} in 2000 (\citealt{Steen03}; \citealt{McK03}) only report flux line values consistent with those that we find in the RGS spectrum. Alternatively one can also estimate the distance of the emission line clouds to the center if one assumes that the gas dynamics are dominated by the central supermassive black hole (e.g. \citealt{KG91}):

\begin{equation}
\label{linedist}
M=\frac{3Rv^2}{2G},
\end{equation}

\noindent where $M$ is the mass of the black hole, $G$ is the constant of gravity, and $v$ is the FWHM is the emission lines. For a mass of $9.8 \pm 2.1 \times 10^6$~$M_\odot$ (\citealt{Den06}) and measured FWHM of the order of 1\,500 $-$ 3\,000~km s$^{-1}$ for the X-ray emission lines, and 2\,500 $-$ 3\,500~km s$^{-1}$ for the UV emission lines, we obtain $R \sim 0.002 - 0.013$~pc. This would locate the emission line clouds well within the BLR of NGC 4593, which is estimated to lie between $3-15$~light-days or $< 0.012$~pc, based on H$\beta$ measurements (\citealt{San95}). Indeed, the estimated high density value $n_e \sim 10^{11}$~cm$^{-3}$ of the gas points to an origin in the BLR and recalls the cases of Mrk 335 (\citealt{Lon08}) and Mrk 841 (\citealt{Lon10}).

\begin{figure*}
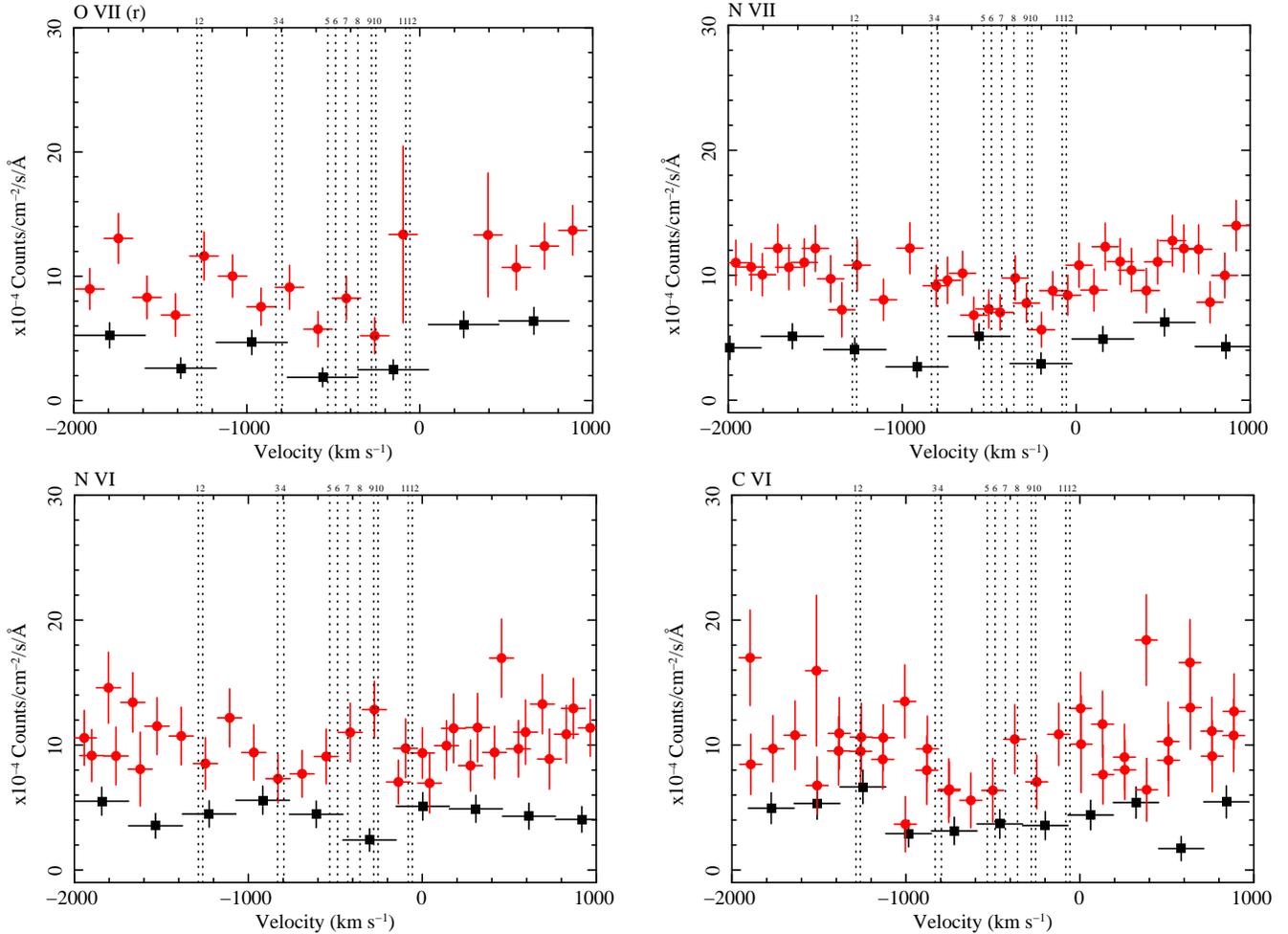

  \hbox{
    \includegraphics[width=6.5cm,angle=-90]{oviir_vspace.ps}
    \includegraphics[width=6.5cm,angle=-90]{nvii_vspace.ps}
  }
  \hbox{
    \includegraphics[width=6.5cm,angle=-90]{nvi_vspace.ps}
    \includegraphics[width=6.5cm,angle=-90]{cvi_vspace.ps}
  }
  \caption{X-ray spectra (LETGS as filled squares; RGS1 and RGS2 as filled circles) in velocity space with respect to the rest-frame of the source at the position of \OVII{}~($r$) (upper left panel), \NVII{}~(upper right panel), \NVI{}~{lower left panel}, and \CVI{}~(lower right panel). For comparison, the kinematic components measured in the UV are shown as dotted vertical lines.}
  \label{xvspace}
\end{figure*}

\subsection{Kinematics of the absorbers}
\label{kinematics}

The results obtained in Sect.~\ref{analysis} show a complex multi-component absorption spectrum in NGC 4593. In the X-rays, three kinematic components (see Table~\ref{WAbest}) are in principle required to obtain a satisfactory fit. In previous X-ray observations (see Sect.~\ref{comparison}) the only reported outflow velocities for a two-zone photoionised gas revealed values consistent with each other (\citealt{Steen03}). The other papers either report a single velocity scenario or no velocity measurement at all. In all cases, the reported outflow velocity values are consistent with each other and they are of the order of $-400$~km s$^{-1}$.

Our {\it slab} fits to the X-ray data allowed for a multi-velocity structure for the absorption system, obtaining 3 kinematic components. In Sect.~\ref{xrayslab} we showed that there is no difference within the uncertainties between the LETGS and RGS kinematic components. Our photoionisation modeling with {\it xabs} also provides a three-velocity structure ($v_{\rm out} \sim -350$, $-460$, and $-1100$~km s$^{-1}$, respectively), which is in agreement within the error bars with those of the {\it slab} modeling. The lower ionisation components $A$ and $B$ share the same outflow velocity at $v_{\rm out} \sim -350$~km s$^{-1}$ which, in turn, differs from that of the intermediate ionisation phase $C$ by less than $2\sigma$. On the other hand, the higher ionisation phase $D$, for which we obtain a higher velocity with {\it xabs} than with {\it slab} albeit with much larger error bars, is outflowing significantly faster than the others, a behaviour often observed in the WA of Seyfert galaxies.

In the UV the kinematic structure is much clearer as 15 distinct kinematic components are seen in \Lya{}, \NV{}, \SiIV{}, and \CIV{} with outflow velocities ranging from $\sim -60$~km s$^{-1}$ to $\sim -1520$~km s$^{-1}$. These kinematic components are unambiguously detected in all the ions, except for \SiIV{} which only shows four troughs belonging to components 1, 2, 9, and 10. The fastest components 13 to 15, with outflow velocities ranging from $-1400$ to $-1520$~km s$^{-1}$ are seen only in \Lya{}. In our UV analysis we also find significant evidence of partial covering in components 3, 5, 6, 10, 11, and 12, a conclusive hallmark of an associated absorber from an outflow, as opposed to a merely intervening cloud.

Whether there is a possible kinematic agreement between the X-ray WA and the UV absorber is difficult to determine because the spectral resolution in the X-ray domain is much poorer and therefore a one-to-one identification is not feasible. To illustrate this, we show in Fig.~\ref{xvspace} the X-ray spectra with LETGS and RGS in velocity space at the position of different ions, with the measured UV kinematic components overplotted.

However, the velocity range of the UV absorbers is roughly similar to that of the X-ray absorber within the errors (i.e. a few hundreds of km s$^{-1}$), thus in principle it is reasonable to assume that some of the UV components might have a counterpart detectable in the X-ray domain. For example, components $A$ and $B$ in the X-rays, which seem to share the same kinematic regime, span a velocity range between $\sim -280$ and $\sim -400$, which would be consistent with UV components 4 and 5. On the other hand, component $C$ in X-rays spreads over an uncertinty range of $\sim -400 - \sim -520$~km s$^{-1}$, which can be kinematically associated to components 6 and/or 7 and, marginally, to component 8 in the UV. Likewise, the fastest X-ray component $D$ ranges between $\sim -900$~and $\sim -1350$~km s$^{-1}$ and is a good match with UV components 11 and 12, and with component 10 at the 2$\sigma$ confidence level. Nevertheless, the X-ray and UV absorbing gas may arise in very different physical environments, so in order to assess whether there is a physical relation between them other properties need to be further examined.

\subsection{Column densities}
\label{densities}

The total ionic column densities of \CIV{} and \NV{} summed over all the UV kinematic components are of the order of $\log N_{\rm CIV} \sim 15.7$~cm$^{-2}$ and $\log N_{\rm NV} \sim 15.8$~cm$^{-2}$, respectively. The bulk of the \NV{} ionic column density is carried by component 9 ($\sim$56\%), and the rest is more or less evenly distributed between components 3 ($\sim$13\%), 6 ($\sim$6\%), 7 ($\sim$10\%), and 10 ($\sim$8\%). For \CIV{}, the ionic column densities are distributed between components 1 ($\sim$70\%) and, to a lower extent, component 9 ($\sim$18\%).

A direct comparison between the X-ray and UV observations, or even among the individual UV components seen by STIS is difficult. As noted in Sect.~\ref{stis}, some of the UV features show a high level of saturation, namely component 1 in \CIV{} and \Lya{}, component 9 in \Lya{}, \NV{}~and \CIV{}, and components 2, 3, 7, 10, and 15 in \Lya{}. In a conservative approach, the column densities reported in Table~\ref{uvslab} should therefore be considered as lower limits. Some general considerations can nevertheless be drawn from these results and, in particular, the comparison with the \XMM{} observation as it was taken quasi-simultaneously with that of \STIS{}. The \NV{} ionic column density measured in the RGS spectrum yields $\log N_{\rm N V} = 15.8 \pm 0.7$~cm$^{-2}$ which, in spite of the large uncertainties, is broadly consistent with the value measured in the UV. The situation is more complicated for \CIV{}, as this feature arises in a very lowly ionised plasma that produces very little absorption in the X-rays. Therefore, only upper limits on its ionic column density can be derived for this ion $\log N_{\rm C IV} < 16.3$~cm$^{-2}$. Since the $N_{\rm C IV}$ measured in the UV does not exceed this value, it can be assumed that there is formally no discrepancy between these results.

The similar outflow velocities and ionic column densities may indicate that the X-ray and UV absorbers originate in the same intervening gas. However, this is at odds with the different ionisation structure depicted by the photoionisation analysis. The ionisation parameter $\xi$ measured in the X-ray WA increases with increasing outflow velocity, while the general trend in the UV is the opposite, with decreasing $\xi$ as the velocity increases (see Table~\ref{uvxabs}). Interestingly, this behaviour is also seen in other Seyfert galaxies such as Mrk 509. For this source \citet{Det11} find 5 distinct ionisation phases with increasing $\xi$ and outflow velocity, whereas the photoionisation analysis of \citet{Arav12}, using \COS{}~and \STIS{}~data, find lower values of the ionisation parameter for the faster troughs. 

\begin{figure}
  \includegraphics[width=8cm,angle=0]{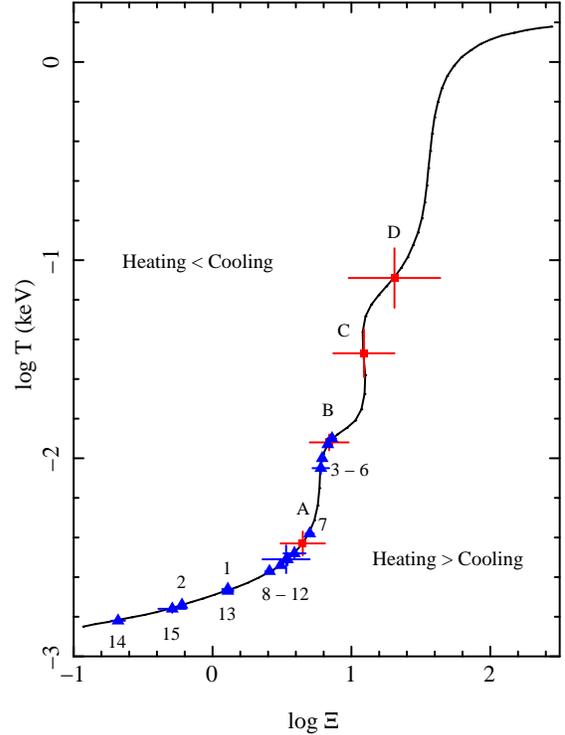}
  \caption{Pressure ionisation parameter as a function of electron temperature (solid line). The X-ray warm absorber components are represented with filled squares ($A$ to $D$). Filled triangles represent the UV absorber components (1 to 15). For clarity, some of the UV labels have been grouped.}
  \label{xitemp}
\end{figure}

\subsection{Stability curve}
\label{scurve}

To further investigate the structure of the X-ray and UV absorbers in NGC 4593 we plot the pressure ionisation parameter $\Xi$ as a function of the electron temperature $T$. The pressure ionisation parameter is defined as $\Xi = L/4 \pi r^2 cp = \xi/4\pi ckT$, where $c$ is the speed of light, $k$ is the constant of Boltzmann, and $T$ is the electron temperature. The different values of $\Xi$ are computed using {\it CLOUDY}, which provided a grid of ionisation parameters $\xi$ and the corresponding electron temperature $T$ for a thin layer of gas illuminated by the ionising continuum. The curve thus generated divides the plane $T-\Xi$ in two regions. Below the curve heating dominates cooling, while above the curve cooling dominates heating (e.g. \citealt{Kro81}). The parts of the curve where $dT/d\Xi < 0$ (i.e. where the curve turns backwards) are unstable against isobaric thermal perturbations (vertical displacements from the curve). The parts of the curve with positive derivative $dT/d\Xi > 0$ define stable regions. Components sharing the same $\Xi$ value on this thermal stability curve are nominally in pressure equilibrium with each other and therefore would likely be part of the same long-lived structure.

In Fig.~\ref{xitemp} we show the thermal stability curve of NGC 4593, with the different X-ray WA and UV absorber components overplotted. The thermal stability curve was generated using the SED at the epoch of the quasi-simultaneous observations of \XMM{} and \STIS{}. Overplotted on the curve are the different phases of the X-ray WA (squares) and UV absorbers (triangles). It can be seen that the X-ray WA components $A$ and $B$ both fall in stable areas of the curve, as well as the higher ionisation component $D$. Component $C$, on the other hand, seems to fall in an unstable branch of the curve. Although WA phases typically fall in stable areas (e.g. Mrk 279, \citealt{Ebr10}), components lying in unstable branches are not uncommon. For example, one of the high-ionisation phases of the WA in Mrk 509 shows this behaviour (\citealt{Det11}), and similarly in NGC 5548 (\citealt{AV10}; Seta et al., in preparation). In general, the majority of WA components are located in thermally stable areas, which would suggest that those components lying in unstable regions would relocate to stable regions on time scales typically short compared to the AGN lifetime.

The UV components are spread over the stable branch of the stability curve. Interestingly, a number of these components tend to cluster around the position of the lower ionistion WA, namely components 7 to 12 around the WA $A$ and components 3 to 6 around the WA $B$. Since they approximately share the same $\Xi$ within the error bars, they are therefore in pressure equilibrium which may indicate that they form part of the same long-lived structure. From this point of view, it is reasonable to assume that the low ionisation WA likely originates in the same intervening gas as the UV absorbers, in particular for the case of the X-ray absorber $B$. The outflow velocity of this WA phase is consistent with those of the UV components 3 to 6, so they would be co-located assuming Keplerian motion. However, in the case of the X-ray absorber $A$, there are kinematic discrepancies that are difficult to reconcile with this scenario as the UV components are outflowing significantly faster.

\subsection{Origin of the absorbers}
\label{origin}

The location of the X-ray and UV absorbers is difficult to estimate. While in principle one could use the definition of the ionisation parameter $\xi$ in Eq.~\ref{xieq} to deduce the distance of the gas to the ionising source $R$, in practice the product $nR^2$ is degenerate. Ideally, variability between observations can be used to determine the density $n$ and therefore put constraints on the distance $R$ (e.g. \citealt{KK95}).

Unfortunately, the distance of the X-ray WA cannot be constrained in this way. For instance, the X-ray WA measured by \citet{Steen03} and in the RGS spectrum in this work has the same best-fit parameters as in our LETGS results within the error bars, even though a change in flux of $\sim$40\% had occured between these observations. Moreover, the column densities of the most relevant ions remained fully consistent within the uncertainties thus preventing us from estimating the distance on grounds of variability. In this respect, because of the lack of monitoring of the WA in NGC 4593 in the last decade, it is difficult to tell whether such changes in the ionising flux happened shortly before our LETGS observation and therefore the gas did not have time to reach the equilibrium again.

There is the possibility to estimate the location of the absorbing gas by other means. For example, \citet{Blu05} used dynamical and momentum conservation arguments to derive an expression that provides an upper limit to the distance. While in principle such an argument can be valid for radiation-driven winds, there is mounting evidence that other mechanisms (e.. shocks, thermal instabilities, magnetic fields) could be involved in the wind launching. In this respect, \citet{Fuk10} studied self-similar solutions for magnetohydrodynamic (MHD) winds in accretion discs, finding that they provide a good match with the observed properties of ionised winds in Seyfert galaxies.

We can derive some rough upper limits on the location of the different components of the ionised winds in NGC 4593 following a very conservative approach. The main assumption is that most of the mass in the WA is concentrated in a layer with thickness $\Delta r$ which is less than or equal to the distance to the central ionising source $R$ so that: $\frac{\Delta r}{R} \leq 1$. The column density along our line of sight can be approximated to $N_{\rm H} \approx n\Delta r$. Combining these conditions with the definition of the ionisation parameter $\xi$ given in Eq.~\ref{xieq}, we are left with

\begin{equation}
\label{Ruplim}
R \leq \frac{L}{N_{\rm H}\xi}.
\end{equation}

For an ionising luminosity of $L = 2.7 \times 10^{43}$~erg s$^{-1}$ at the epoch of the \XMM{} and \STIS{} observations, we find that the lower ionisation components $A$ and $B$ are located within $R_{\rm A} \leq 2.9$~kpc and $R_{\rm B} \leq 270$~pc, respectively, from the central engine. The location of the WA component $C$ is more constrained within $R_{\rm C} \leq 15$~pc. For the highest ionisation component $D$, given the uncertainty in the determination of its column density, we obtain an upper limit to the distance $R_{\rm D}$ ranging between $6$ and $29$~pc. Following the same reasoning the UV absorbers are located at kpc distances.

Although these limits are rather loose, they clearly indicate that our line of sight is crossing several clouds of gas located at different distances from the central engine, with decreasing ionisation state as the distance increases. This scenario is sketched by \citet{Kaz12} in their Fig. 8, where a line of sight at a given inclination angle intersects different parts of a stratified wind, each with different $\xi$, column density and velocity. While it is still possible that some of the UV features originate in the same gas that causes the low-ionisation X-ray absorption, the model depicted by \citet{Kaz12} suggest that the variety of ionisation phases seen in the gas belong to different strata originally launched from different parts of the disc.

\subsection{Energetics and feedback implications}
\label{energetics}

We can estimate how much mass is carried out of the central regions of NGC 4593 by the outflows and assess whether it may have any impact on the surrounding ISM of the host galaxy. If the outflow is in the form of a partial thin spherical shell at a distance $R$ from the central source moving radially at constant speed $v$, the mass outflow rate is given by

\begin{equation}
\label{masseq}
\dot{M}_{\rm out} = \mu m_{\rm p}vnR^2\Omega = \mu m_{\rm p}v\frac{L}{\xi}\Omega,
\end{equation}

\noindent where $\mu = 1.4$ is the mean atomic mass per proton, $m_{\rm p}$ is the proton mass, and $\Omega$ is the solid angle subtended by the outflow (between 0 and $4\pi$~sr). The main source of uncertainty in the determination of the mass outflow rate is introduced by the parameter $\Omega$, which is typically assumed to be $\Omega = \pi/2$~sr, inferred from the observed type-1/type2 ratio in nearby Seyfert galaxies (\citealt{MR95}) and the fact that ionised outflows are seen in approximately half of the observed Seyfert 1 galaxies (\citealt{Dunn07}). We note that this value may not hold if the outflows are bent because of gravitational and/or magnetic processes.

In this way we obtain $\dot{M}_{\rm A} \simeq 3.7\Omega$~$M_\odot$~yr$^{-1}$, $\dot{M}_{\rm B} \simeq 0.68\Omega$~$M_\odot$~yr$^{-1}$, $\dot{M}_{\rm C} \simeq 0.18\Omega$~$M_\odot$~yr$^{-1}$, and $\dot{M}_{\rm D} \simeq 0.11\Omega$~$M_\odot$~yr$^{-1}$ for the WA components $A$ to $D$, respectively. We expressed these mass outflow rates as a function of $\Omega$ in order to reflect the uncertainty associated to this parameter.

For comparison, the mass accretion rate is obtained from the bolometric luminosity using

\begin{equation}
\label{accrate}
\dot{M}_{\rm acc} = \frac{L_{\rm bol}}{\eta c^2}.
\end{equation}

\noindent The bolometric luminosities of NGC 4593 derived from the SED is $L_{\rm bol} = 3.7 \times 10^{43}$~erg s$^{-1}$, and $L_{\rm bol} = 8.15 \times 10^{43}$~erg s$^{-1}$ for the LETGS and RGS observations, respectively. Assuming a nominal accretion efficiency of $\eta = 0.1$, the mass accretion rate results in $\dot{M}_{\rm acc} = 6.5 \times 10^{-3}$~$M_\odot$~yr$^{-1}$, and $\dot{M}_{\rm acc} = 1.44 \times 10^{-2}$~$M_\odot$~yr$^{-1}$, respectively. One can see that the accretion rate at the epoch of the RGS observation almost doubled that of the LETGS observation, assuming that the accretion efficiency remained the same. The total mass outflow rate of the X-ray WA is therefore higher than the mass accretion rate, as often seen in other sources (e.g. \citealt{Cos10}; \citealt{CK12}). As an example, assuming that the absorbers subtend a solid angle $\Omega = \pi/2$~sr, the mass outflow rates of the X-ray components would be $\dot{M}_{\rm A} \simeq 5.9$~$M_\odot$~yr$^{-1}$, $\dot{M}_{\rm B} \simeq 1.1$~$M_\odot$~yr$^{-1}$, $\dot{M}_{\rm C} \simeq 0.3$~$M_\odot$~yr$^{-1}$, and $\dot{M}_{\rm D} \simeq 0.2$~$M_\odot$~yr$^{-1}$.

The lowest ionisation component $A$ is located at $\sim$kpc distances from the central source. Therefore, this outflow likely originated in an earlier episode of activity of NGC 4593 (e.g. \citealt{King11}) and it is unrelated to the current AGN emission detected in these observations. In a conservative approach, we will not consider this component in the next energetics estimations. Thus, the total mass outflow rate summed over the remaining components is $\dot{M}_{\rm out} \sim 1.6$~$M_\odot$~yr$^{-1}$, between $\sim 100$ and $\sim 250$ times the mass accretion rate at the epoch of the RGS and LETGS observations.

The total kinetic energy per unit time carried by the outflow is the sum of the kinetic luminosity injected into the medium by the different components, and is defined as $L_{\rm KE} = 1/2 \dot{M}_{\rm out}v^2$. Excluding component $A$, we find it to be $L_{\rm KE} \simeq 7.8 \times 10^{40}\Omega$~erg s$^{-1}$, expressed as a function of $\Omega$. The total kinetic energy injected in the medium represents a small fraction of the bolometric luminosity of the source: $L_{\rm KE}/L_{\rm bol} \sim 0.1\Omega$\%, for $L_{\rm bol} = 8.15 \times 10^{43}$~erg s$^{-1}$. For a typical value $\Omega = \pi/2$~sr, we thus obtain $L_{\rm KE}/L_{\rm bol} \sim 0.15$\%. These values are in agreement with those estimated for other nearby Seyfert galaxies reported in \citet{CK12}.

Overall, the outflows in NGC 4593 seem to be too weak to produce a significant impact in the surrounding ISM. Many AGN feedback models require that a significant fraction of the AGN radiative output ($\sim$5\%) has to be fed back in the form of kinetic and internal motion energy in outflows in order to reproduce the observed $M-\sigma$ relation (e.g. \citealt{SR98}; \citealt{WL03}), at least one order of magnitude above the values we measure for NGC 4593.

However, if we consider that the measured kinetic luminosity is injected into the ISM at this rate throughout the whole AGN lifetime, which can be up to $t_{\rm AGN} \sim 10^{8-9}$~yr (\citealt{Gilli09}; \citealt{Ebr09}), the total kinetic energy fed back to the medium would be $E_{\rm KE} = L_{\rm KE}t_{\rm AGN} \sim 10^{56-57}$~erg. This value is roughly consistent with the energy required to disrupt the ISM of a typical galaxy, which is estimated to be $\sim 10^{57}$~erg (\citealt{Kron10}). A more conservative approach may consider that the expansion of the outflowing gas, with a corresponding decrease in density as it dilutes in the ISM, could increase the cross-section of the material thus favouring the momentum coupling from absorption of the ionising flux. In this case the required feedback to alter the ISM and supress star formation in the host galaxy might be reduced by about one order of magnitude to $\sim 0.5$\% of the luminosity (\citealt{HE10}). While our estimated kinetic luminosity is still lower than this value, it is therefore very unlikely that the outflows seen in NGC 4593 can significantly affect the medium of the host galaxy in a single AGN episode. However, if we integrate these contributions over the entire lifetime of the AGN, it results that the outflows (even for a low luminosity AGN hosting a relatively small SMBH like NGC 4593) may have a small (but not negligible) influence on the surroundings of the host galaxy.

\section{Conclusions}
\label{conclusions}

We have analysed the soft X-ray and UV spectra of the Seyfert 1 galaxy NGC 4593 obtained with \Chan{}-LETGS, \XMM{} RGS, and HST/STIS, respectively. In the X-rays there is significant evidence of a multi-phase WA with four distinct degrees of ionisation: $\log \xi = 1.0$, $\log \xi = 1.7$, $\log \xi = 2.4$, and $\log \xi = 3.0$, respectively. These components are outflowing, being the low-ionisation components $A$ and $B$ slower ($v_{\rm out} \sim -350$~km s$^{-1}$) than the high-ionisation components $C$ ($v_{\rm out} = -460$~km s$^{-1}$), and $D$ ($v_{\rm out} \sim -1100$~km s$^{-1}$).

In the UV, we detected a wealth of absorption troughs in the \Lya{}, \NV{}, \SiIV{}, and \CIV{} regions. These UV absorbers are also blueshifted with respect to the systemic velocity of the source and therefore are also outflowing. They can be attributed to 15 distinct kinematic components ranging between $v_{\rm out} \sim -60$~km s$^{-1}$ and $v_{\rm out} \sim -1520$~km s$^{-1}$. Attending to the ionisation state of the gas, UV components 7 to 12 might be the counterparts of the low-ionisation phase $A$, whereas components 3 to 6 show similar values as component $B$. Components $C$ and $D$ are probably too ionised to produce any significant absorption in the UV. Furthermore, the X-ray low-ionisation WA is in pressure equilibrium with the UV absorber thus indicating that they may form part of the same long-lived structure. The predicted \NV{} and \CIV{} (only upper limits in this case) column densities from the X-rays are consistent with those measured in the UV. There are, however, kinematic discrepancies that may prevent both absorbers from originating in the same intervening gas.

We put some constraints on the location of the WA assuming that its mass is concentrated in a layer with thickness lower than or equal to the distance to central source. We derive an upper limit for the distance of high-ionisation phases to the central ionising source of $\lesssim 6 - 29$~pc, while the low-ionisation phases are located at hundreds of pc. This is consistent with a scenario in which our line of sight intersects several parts of a stratified wind, each with different ionisation state, density, and velocity, as sketched in \citet{Kaz12}.

The total kinetic luminosity of the outflows injected into the ISM of NGC 4593 represents only a tiny fraction of the bolometric luminosity, $L_{\rm KE}/L_{\rm bol} \sim 0.1\Omega$\%, where $\Omega$ is the solid angle subtended by the outflow. This is too low to produce any significant impact in the medium of the host galaxy on a single episode of activity, but it may produce small (but not negligible) effects when integrated over the whole AGN lifetime.

\section*{Acknowledgments}

SRON is supported financially by NWO, the Netherlands Organisation for Scientific Research. This work is based on observations obtained with XMM-Newton, an ESA science mission with instruments and contributions directly funded by ESA Member States and NASA. The scientific results reported in this article are based on observations made by the Chandra X-ray Observatory. Some of the data presented in this paper were obtained from the Mikulski Archive for Space Telescopes (MAST). STScI is operated by the Association of Universities for Research in Astronomy, Inc., under NASA contract NAS5-26555. We thank the anonymous referee for his/her comments.

\bsp

\label{lastpage}

\end{document}